\def\includeeventviews{}
\ifx \sciencesubmit\undefined
\documentclass[aps,prd,reprint,superscriptaddress,twocolumn]{revtex4-1}
\usepackage{hyperref}
\else
\documentclass[12pt]{article}
\usepackage{times}
\renewenvironment{abstract}{%
\begin{quote} \bf}
{\end{quote}}

\newcommand{\affiliation}[1]{}
\newcommand{\collaboration}[1]{}
\newcommand{\noaffiliation}{}

\newenvironment{acknowledgments}{\section*{Acknowledgments}}{}
\usepackage{scicite}

\usepackage[nolists,nomarkers]{endfloat}
\fi

\usepackage{graphicx}
\usepackage{amsmath}
\usepackage{amssymb}
\usepackage{url}
\usepackage{units}
\usepackage{color}

\def\apj{Astrophysical Journal }

\def\prd{Physical Review D }

\ifx \sciencesubmit\undefined
\begin{document}
\fi

\title{Evidence for High-Energy Extraterrestrial Neutrinos at the IceCube Detector}

\ifx \sciencesubmit\undefined
\affiliation{III. Physikalisches Institut, RWTH Aachen University, D-52056 Aachen, Germany}
\affiliation{School of Chemistry \& Physics, University of Adelaide, Adelaide SA, 5005 Australia}
\affiliation{Dept.~of Physics and Astronomy, University of Alaska Anchorage, 3211 Providence Dr., Anchorage, AK 99508, USA}
\affiliation{CTSPS, Clark-Atlanta University, Atlanta, GA 30314, USA}
\affiliation{School of Physics and Center for Relativistic Astrophysics, Georgia Institute of Technology, Atlanta, GA 30332, USA}
\affiliation{Dept.~of Physics, Southern University, Baton Rouge, LA 70813, USA}
\affiliation{Dept.~of Physics, University of California, Berkeley, CA 94720, USA}
\affiliation{Lawrence Berkeley National Laboratory, Berkeley, CA 94720, USA}
\affiliation{Institut f\"ur Physik, Humboldt-Universit\"at zu Berlin, D-12489 Berlin, Germany}
\affiliation{Fakult\"at f\"ur Physik \& Astronomie, Ruhr-Universit\"at Bochum, D-44780 Bochum, Germany}
\affiliation{Physikalisches Institut, Universit\"at Bonn, Nussallee 12, D-53115 Bonn, Germany}
\affiliation{Universit\'e Libre de Bruxelles, Science Faculty CP230, B-1050 Brussels, Belgium}
\affiliation{Vrije Universiteit Brussel, Dienst ELEM, B-1050 Brussels, Belgium}
\affiliation{Dept.~of Physics, Chiba University, Chiba 263-8522, Japan}
\affiliation{Dept.~of Physics and Astronomy, University of Canterbury, Private Bag 4800, Christchurch, New Zealand}
\affiliation{Dept.~of Physics, University of Maryland, College Park, MD 20742, USA}
\affiliation{Dept.~of Physics and Center for Cosmology and Astro-Particle Physics, Ohio State University, Columbus, OH 43210, USA}
\affiliation{Dept.~of Astronomy, Ohio State University, Columbus, OH 43210, USA}
\affiliation{Dept.~of Physics, TU Dortmund University, D-44221 Dortmund, Germany}
\affiliation{Dept.~of Physics, University of Alberta, Edmonton, Alberta, Canada T6G 2E1}
\affiliation{D\'epartement de physique nucl\'eaire et corpusculaire, Universit\'e de Gen\`eve, CH-1211 Gen\`eve, Switzerland}
\affiliation{Dept.~of Physics and Astronomy, University of Gent, B-9000 Gent, Belgium}
\affiliation{Dept.~of Physics and Astronomy, University of California, Irvine, CA 92697, USA}
\affiliation{Laboratory for High Energy Physics, \'Ecole Polytechnique F\'ed\'erale, CH-1015 Lausanne, Switzerland}
\affiliation{Dept.~of Physics and Astronomy, University of Kansas, Lawrence, KS 66045, USA}
\affiliation{Dept.~of Astronomy, University of Wisconsin, Madison, WI 53706, USA}
\affiliation{Dept.~of Physics and Wisconsin IceCube Particle Astrophysics Center, University of Wisconsin, Madison, WI 53706, USA}
\affiliation{Institute of Physics, University of Mainz, Staudinger Weg 7, D-55099 Mainz, Germany}
\affiliation{Universit\'e de Mons, 7000 Mons, Belgium}
\affiliation{T.U. Munich, D-85748 Garching, Germany}
\affiliation{Bartol Research Institute and Department of Physics and Astronomy, University of Delaware, Newark, DE 19716, USA}
\affiliation{Dept.~of Physics, University of Oxford, 1 Keble Road, Oxford OX1 3NP, UK}
\affiliation{Dept.~of Physics, University of Wisconsin, River Falls, WI 54022, USA}
\affiliation{Oskar Klein Centre and Dept.~of Physics, Stockholm University, SE-10691 Stockholm, Sweden}
\affiliation{Department of Physics and Astronomy, Stony Brook University, Stony Brook, NY 11794-3800, USA}
\affiliation{Department of Physics, Sungkyunkwan University, Suwon 440-746, Korea}
\affiliation{Dept.~of Physics and Astronomy, University of Alabama, Tuscaloosa, AL 35487, USA}
\affiliation{Dept.~of Astronomy and Astrophysics, Pennsylvania State University, University Park, PA 16802, USA}
\affiliation{Dept.~of Physics, Pennsylvania State University, University Park, PA 16802, USA}
\affiliation{Dept.~of Physics and Astronomy, Uppsala University, Box 516, S-75120 Uppsala, Sweden}
\affiliation{Dept.~of Physics, University of Wuppertal, D-42119 Wuppertal, Germany}
\affiliation{DESY, D-15735 Zeuthen, Germany}

\author{M.~G.~Aartsen}
\affiliation{School of Chemistry \& Physics, University of Adelaide, Adelaide SA, 5005 Australia}
\author{R.~Abbasi}
\affiliation{Dept.~of Physics and Wisconsin IceCube Particle Astrophysics Center, University of Wisconsin, Madison, WI 53706, USA}
\author{Y.~Abdou}
\affiliation{Dept.~of Physics and Astronomy, University of Gent, B-9000 Gent, Belgium}
\author{M.~Ackermann}
\affiliation{DESY, D-15735 Zeuthen, Germany}
\author{J.~Adams}
\affiliation{Dept.~of Physics and Astronomy, University of Canterbury, Private Bag 4800, Christchurch, New Zealand}
\author{J.~A.~Aguilar}
\affiliation{D\'epartement de physique nucl\'eaire et corpusculaire, Universit\'e de Gen\`eve, CH-1211 Gen\`eve, Switzerland}
\author{M.~Ahlers}
\affiliation{Dept.~of Physics and Wisconsin IceCube Particle Astrophysics Center, University of Wisconsin, Madison, WI 53706, USA}
\author{D.~Altmann}
\affiliation{Institut f\"ur Physik, Humboldt-Universit\"at zu Berlin, D-12489 Berlin, Germany}
\author{J.~Auffenberg}
\affiliation{Dept.~of Physics and Wisconsin IceCube Particle Astrophysics Center, University of Wisconsin, Madison, WI 53706, USA}
\author{X.~Bai}
\thanks{Physics Department, South Dakota School of Mines and Technology, Rapid City, SD 57701, USA}
\affiliation{Bartol Research Institute and Department of Physics and Astronomy, University of Delaware, Newark, DE 19716, USA}
% ...

\collaboration{IceCube Collaboration}
\noaffiliation
\else
\author{IceCube Collaboration \\
\normalsize{Corresponding authors:}\\
\normalsize{E-mail: ckopper@icecube.wisc.edu (C.K.); naoko@icecube.wisc.edu (N.K.);}\\
\normalsize{nwhitehorn@icecube.wisc.edu (N.W.)}\\
}
\date{}

\begin{document}
\maketitle
\fi

\begin{abstract}
We report on results of an all-sky search for high-energy neutrino events interacting within the IceCube neutrino detector conducted between May 2010 and May 2012.
The search follows up on the previous detection of two PeV neutrino events, with improved sensitivity and extended energy coverage down to approximately 30 TeV.
Twenty-six additional events were observed, substantially more than expected from atmospheric backgrounds.
Combined, both searches reject a purely atmospheric origin for the twenty-eight events at the $4\sigma$ level.
These twenty-eight events, which include the highest energy neutrinos ever observed, have flavors, directions, and energies inconsistent with those expected from the atmospheric muon and neutrino backgrounds.
These properties are, however, consistent with generic predictions for an additional component of extraterrestrial origin.
\end{abstract}

\ifx \sciencesubmit\undefined
\maketitle
\else
\fi

\begin{figure}
\includegraphics[width=0.465\linewidth]{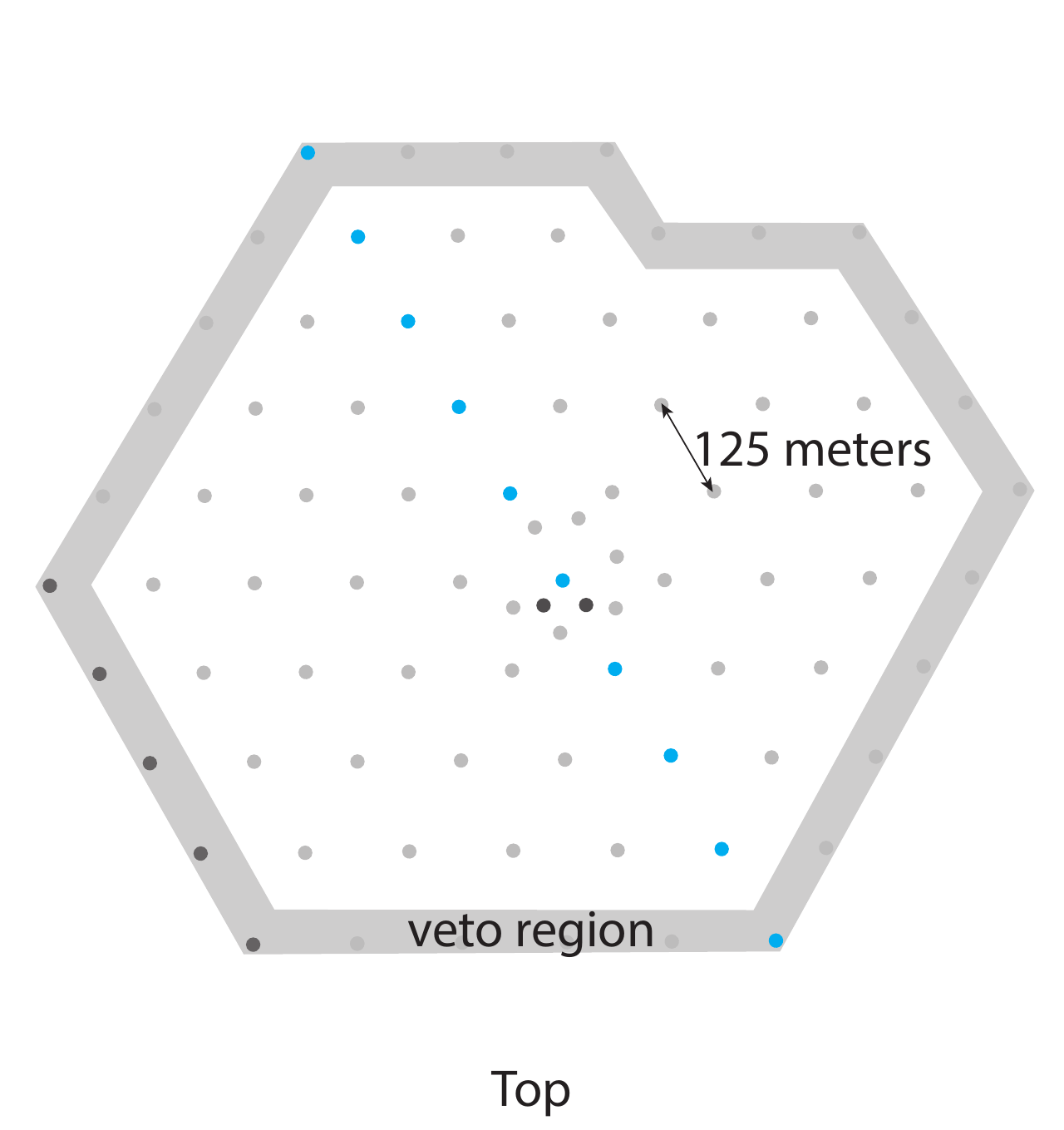}
\includegraphics[width=0.515\linewidth]{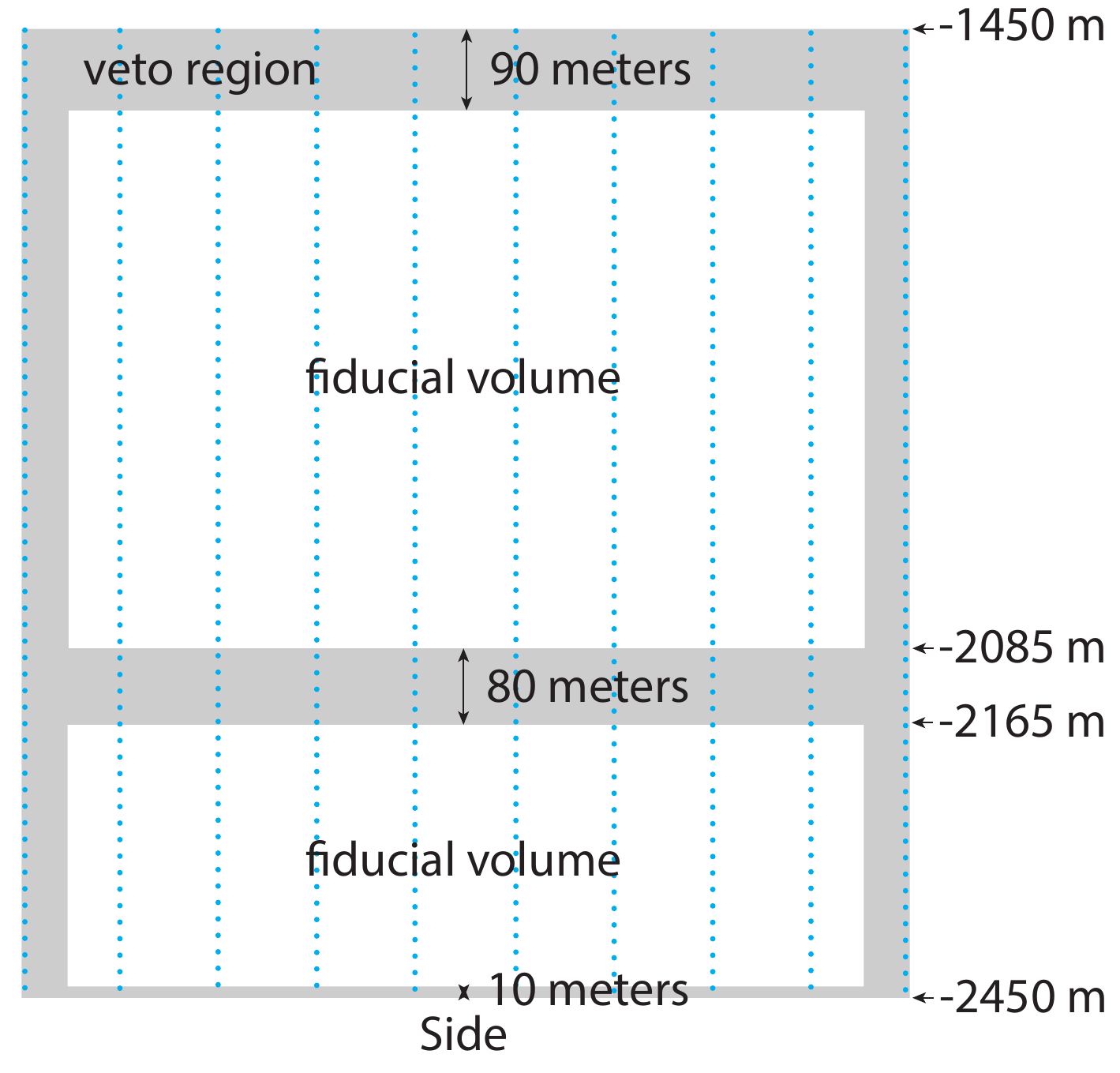}
\caption{%
Drawing of the IceCube array.
Results here are from the complete pictured detector for 2011-2012 and from a partial detector missing the dark gray strings in the bottom left corner for the 2010-2011 season.
The side view (right) shows a cross-section of the detector indicated in the top view (left) in blue.
Events producing first light in the veto region (shaded area) were discarded as entering tracks (usually from cosmic ray muons entering the detector).
Most background events are nearly vertical, requiring a thick veto cap at the top of the detector.
The shaded region in the middle contains ice of high dust concentration \cite{spice}.
Because of the high degree of light absorption in this region, near horizontal events could have entered here without being tagged at the sides of the detector without a dedicated tagging region.
}
\label{fig:icecube_veto}
\end{figure}

\section*{Introduction}

High-energy neutrino observations can provide insight into the long-standing problem of the origins and acceleration mechanisms of high-energy cosmic rays.
As cosmic ray protons and nuclei are accelerated, they interact with gas and background light to produce charged pions and kaons which then decay, emitting neutrinos with energies proportional to the energies of the high-energy protons that produced them.
These neutrinos can be detected on Earth in large underground detectors by the production of secondary leptons and hadronic showers when they interact with the detector material.
IceCube, a large-volume Cherenkov detector \cite{2006APh....26..155I} made of 5160 photomultipliers (PMTs) at depths between 1450 and 2450 meters in natural Antarctic ice (Fig.~\ref{fig:icecube_veto}), has been designed to detect these neutrinos at TeV-PeV energies.
Recently, the Fermi collaboration presented evidence for acceleration of low energy (GeV) cosmic-ray protons in supernova remnants \cite{FermiPi0}; neutrino observations with IceCube would probe sources of cosmic rays at far higher energies.

A recent IceCube search for neutrinos of EeV ($10^6$ TeV) energy found two events at energies of 1 PeV ($10^3$ TeV), above what is generally expected from atmospheric backgrounds and a possible hint of an extraterrestrial source \cite{ehepaper}.
Although that analysis had some sensitivity to neutrino events of all flavors above 1 PeV, it was most sensitive to $\nu_\mu$ events above 10 PeV from the region around the horizon, above which the energy threshold increased sharply to 100 PeV.
As a result, it had only limited sensitivity to the type of events found, which were typical of either $\nu_e$ or neutral current events and at the bottom of the detectable energy range, preventing a detailed understanding of the population from which they arose and an answer to the question of their origin.

Here we present a follow-up analysis designed to characterize the flux responsible for these events by conducting an exploratory search for neutrinos at lower energies with interaction vertices well contained within the detector volume, discarding events containing muon tracks originating outside of IceCube (Fig.~\ref{fig:icecube_veto}).
This event selection (see Materials and Methods) allows the resulting search to have approximately equal sensitivity to neutrinos of all flavors and from all directions.
We obtained nearly full efficiency for interacting neutrinos above several hundred TeV, with some sensitivity extending to neutrino energies as low as 30 TeV; see Fig.~\ref{fig:effective_area} in Materials and Methods.
The data-taking period is shared with the earlier high-energy analysis: data shown were taken during the first season running with the completed IceCube array (86~strings, between May 2011 and May 2012) and the preceding construction season (79~strings, between May 2010 and May 2011), with a total combined live time of 662 days.

\section*{Results}
\label{sec:results}

\begin{figure}
\includegraphics[width=\linewidth]{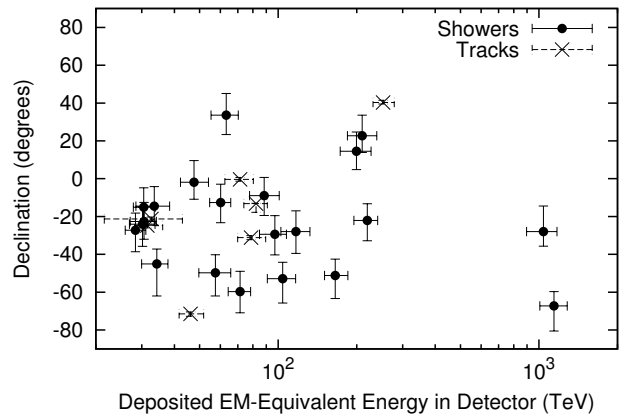}
\caption{%
Distribution of best-fit deposited energies and declinations.
Seven of the events contain muons (crosses) with an angular resolution of about $1^\circ$, while the remainder are either electromagnetic or hadronic showers (filled circles) with an energy-dependent resolution of about $15^\circ$.
Error bars are 68\% confidence intervals including both statistical and systematic uncertainties.
Energies shown are the energy deposited in the detector assuming all light emission is from electromagnetic showers.
For $\nu_e$ charged-current events this equals the neutrino energy; otherwise it is a lower limit on the neutrino energy.
The gap in $E_{dep}$ between 300 TeV and 1 PeV does not appear to be significant: gaps of this size or larger appear in 28\% of realizations of the best-fit continuous power-law flux.
}
\label{fig:energyzenith}
\end{figure}

\begin{figure}
\includegraphics[width=\linewidth]{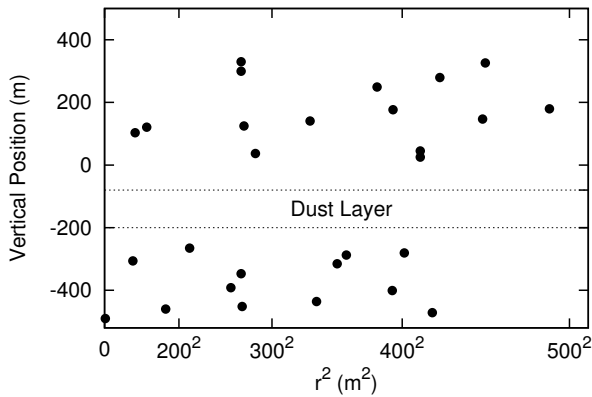}
\caption{%
Coordinates of the first detected light from each event in the final sample.
Penetrating muon events are first detected predominantly at the detector boundaries (top and right sides) where they first make light after crossing the veto layer.
Neutrino events should interact uniformly throughout the approximately cylindrical detector volume, forming a uniform distribution in $(r^2, z)$ with the exception of interactions in the less-transparent ice region marked ``dust layer'', which is treated as part of the detector boundary for purposes of our event selection.
The observed events are consistent with a uniform distribution.
}
\label{fig:firstlightpositions}
\end{figure}

In the two-year dataset, 28 events with in-detector deposited energies between 30 and 1200 TeV were observed (Fig.~\ref{fig:energyzenith}, Table~\ref{tab:events}) on an expected background of $10.6^{+5.0}_{-3.6}$ events from atmospheric muons and neutrinos; see Materials and Methods.
The two most energetic of these were the previously reported PeV events \cite{ehepaper}.
Seven events contained clearly identifiable muon tracks, whereas the remaining twenty-one were shower-like, consistent with neutrino interactions other than $\nu_\mu$ charged-current.
Four of the low energy track-like events started near the detector boundary and are downgoing, consistent with the properties of the expected $6.0 \pm 3.4$ background atmospheric muons, as measured from a control sample of penetrating muons in data.
One of these---the only such event in the sample---had hits in the IceTop surface air shower array compatible with its arrival time and direction in IceCube (event 28).
The points at which the remaining events were first observed were uniformly distributed throughout the detector (Fig.~\ref{fig:firstlightpositions}).
This is consistent with expectations for neutrino events and inconsistent with backgrounds from penetrating muons or with detector artifacts, which would have been expected to trace the locations of either the fiducial volume boundary or the positions of the instrumentation.

\begin{table}
\begin{tabular}{c|c|c|c|c|c|c}
   & Dep. Energy & Time  & Decl. & R.A. & Med. Angular    &   Event   \\
ID & (TeV)            & (MJD) & (deg.)      & (deg.)          & Error (deg.) & Type\\
\hline
1 & $47.6 \,^{+6.5}_{-5.4}$ & 55351 & $-1.8$ & $35.2$ & $16.3$ & Shower \\
2 & $117 \,^{+15}_{-15}$ & 55351 & $-28.0$ & $282.6$ & $25.4$ & Shower \\
3 & $78.7 \,^{+10.8}_{-8.7}$ & 55451 & $-31.2$ & $127.9$ & $\lesssim 1.4$ & Track \\
4 & $165 \,^{+20}_{-15}$ & 55477 & $-51.2$ & $169.5$ & $7.1$ & Shower \\
5 & $71.4 \,^{+9.0}_{-9.0}$ & 55513 & $-0.4$ & $110.6$ & $\lesssim 1.2$ & Track \\
6 & $28.4 \,^{+2.7}_{-2.5}$ & 55568 & $-27.2$ & $133.9$ & $9.8$ & Shower \\
7 & $34.3 \,^{+3.5}_{-4.3}$ & 55571 & $-45.1$ & $15.6$ & $24.1$ & Shower \\
8 & $32.6 \,^{+10.3}_{-11.1}$ & 55609 & $-21.2$ & $182.4$ & $\lesssim 1.3$ & Track \\
9 & $63.2 \,^{+7.1}_{-8.0}$ & 55686 & $33.6$ & $151.3$ & $16.5$ & Shower \\
10 & $97.2 \,^{+10.4}_{-12.4}$ & 55695 & $-29.4$ & $5.0$ & $8.1$ & Shower \\
11 & $88.4 \,^{+12.5}_{-10.7}$ & 55715 & $-8.9$ & $155.3$ & $16.7$ & Shower \\
12 & $104 \,^{+13}_{-13}$ & 55739 & $-52.8$ & $296.1$ & $9.8$ & Shower \\
13 & $253 \,^{+26}_{-22}$ & 55756 & $40.3$ & $67.9$ & $\lesssim 1.2$ & Track \\
14 & $1041 \,^{+132}_{-144}$ & 55783 & $-27.9$ & $265.6$ & $13.2$ & Shower \\
15 & $57.5 \,^{+8.3}_{-7.8}$ & 55783 & $-49.7$ & $287.3$ & $19.7$ & Shower \\
16 & $30.6 \,^{+3.6}_{-3.5}$ & 55799 & $-22.6$ & $192.1$ & $19.4$ & Shower \\
17 & $200 \,^{+27}_{-27}$ & 55800 & $14.5$ & $247.4$ & $11.6$ & Shower \\
18 & $31.5 \,^{+4.6}_{-3.3}$ & 55924 & $-24.8$ & $345.6$ & $\lesssim 1.3$ & Track \\
19 & $71.5 \,^{+7.0}_{-7.2}$ & 55926 & $-59.7$ & $76.9$ & $9.7$ & Shower \\
20 & $1141 \,^{+143}_{-133}$ & 55929 & $-67.2$ & $38.3$ & $10.7$ & Shower \\
21 & $30.2 \,^{+3.5}_{-3.3}$ & 55937 & $-24.0$ & $9.0$ & $20.9$ & Shower \\
22 & $220 \,^{+21}_{-24}$ & 55942 & $-22.1$ & $293.7$ & $12.1$ & Shower \\
23 & $82.2 \,^{+8.6}_{-8.4}$ & 55950 & $-13.2$ & $208.7$ & $\lesssim 1.9$ & Track \\
24 & $30.5 \,^{+3.2}_{-2.6}$ & 55951 & $-15.1$ & $282.2$ & $15.5$ & Shower \\
25 & $33.5 \,^{+4.9}_{-5.0}$ & 55967 & $-14.5$ & $286.0$ & $46.3$ & Shower \\
26 & $210 \,^{+29}_{-26}$ & 55979 & $22.7$ & $143.4$ & $11.8$ & Shower \\
27 & $60.2 \,^{+5.6}_{-5.6}$ & 56009 & $-12.6$ & $121.7$ & $6.6$ & Shower \\
28 & $46.1 \,^{+5.7}_{-4.4}$ & 56049 & $-71.5$ & $164.8$ & $\lesssim 1.3$ & Track \\
\end{tabular}
\caption{%
Properties of the 28 events.
Shown are the deposited electromagnetic-equivalent energy (the energy deposited by the events in IceCube assuming all light was made in electromagnetic showers) as well as the arrival time and direction of each event and its topology (track or shower-like).
The energy shown is equal to the neutrino energy for $\nu_e$ charged-current events, within experimental uncertainties, and is otherwise a lower limit on the neutrino energy due to exiting muons or neutrinos.
Errors on energy and the angle include both statistical and systematic effects.
Systematic uncertainties on directions for shower-like events were determined on an individual basis; track systematic uncertainties here are equal to $1^\circ$, which is an upper limit from studies of the cosmic ray shadow of the moon \cite{moonpaper}.
}
\label{tab:events}
\end{table}

As part of our blind analysis, we tested a pre-defined fixed atmospheric-only neutrino flux model~\cite{Honda2006} including a benchmark charm component~\cite{EnbergPrompt}, reevaluated using current measurements of the cosmic-ray spectrum in this energy range~\cite{GaisserImprovedCRSpectrum, IC59DiffuseNOWPaper}.
This adds an additional 1.5 charm neutrinos to our mean background estimate and predicts on average 6.1 ($\pi/K$ and charm) background neutrinos on top of the $6.0 \pm 3.4$ background muon events.
Significance was evaluated based on the number of events, the total collected photomultiplier charge of each, and the events' reconstructed energies and directions (see Materials and Methods).
Our procedure does not allow us to separately incorporate uncertainties on the various background components.
To nevertheless obtain an indication of the range of possible significances we have calculated values relative to background-only hypotheses with charm at the level called ``standard'' in \cite{EnbergPrompt} as a benchmark flux as well as at the level of our current 90\% CL experimental bounds \cite{IC59DiffuseNOWPaper} (corresponding to 3.8 times ``standard'').
To prevent possible confirmation bias, we split the data set into two samples.
For the 26 new events reported here, using the benchmark flux, we obtain a significance of $3.3\sigma$ (one-sided).
Combined by Fisher's method with the $2.8\sigma$ observation of the earlier analysis where the two highest energy events were originally reported \cite{ehepaper}, and which uses the same benchmark atmospheric neutrino flux model, we obtain a final significance for the entire data set of 28 events of $4.1\sigma$.
The same calculation performed a posteriori on all 28 events gives $4.8\sigma$.
These two final significances would be reduced to $3.6\sigma$ and $4.5\sigma$, respectively, using charm at the level of our current 90\% CL experimental bound.


\section*{Discussion}
\label{sec:interpretation}

\begin{figure*}
\includegraphics[width=0.48\linewidth]{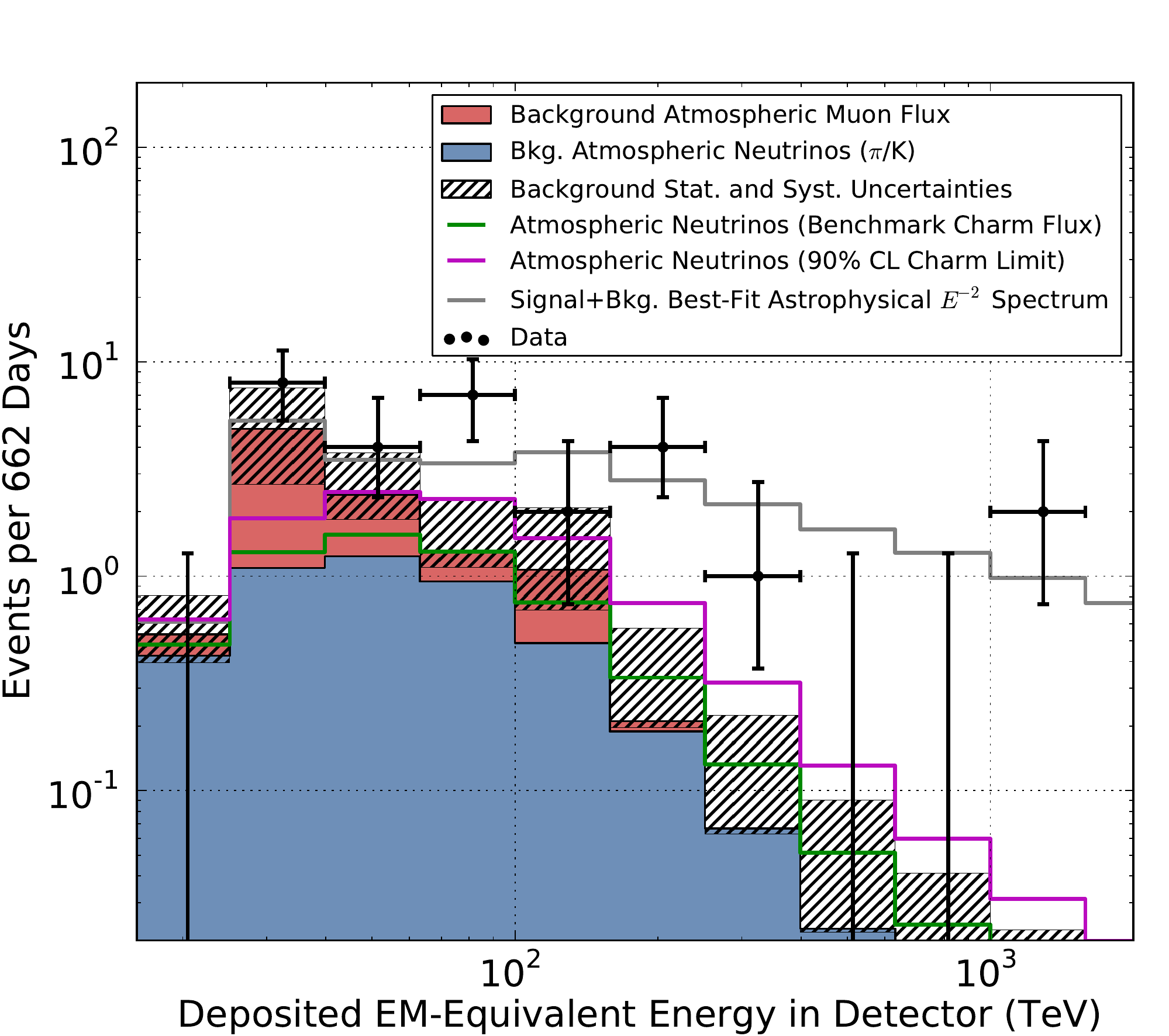}
\includegraphics[width=0.48\linewidth]{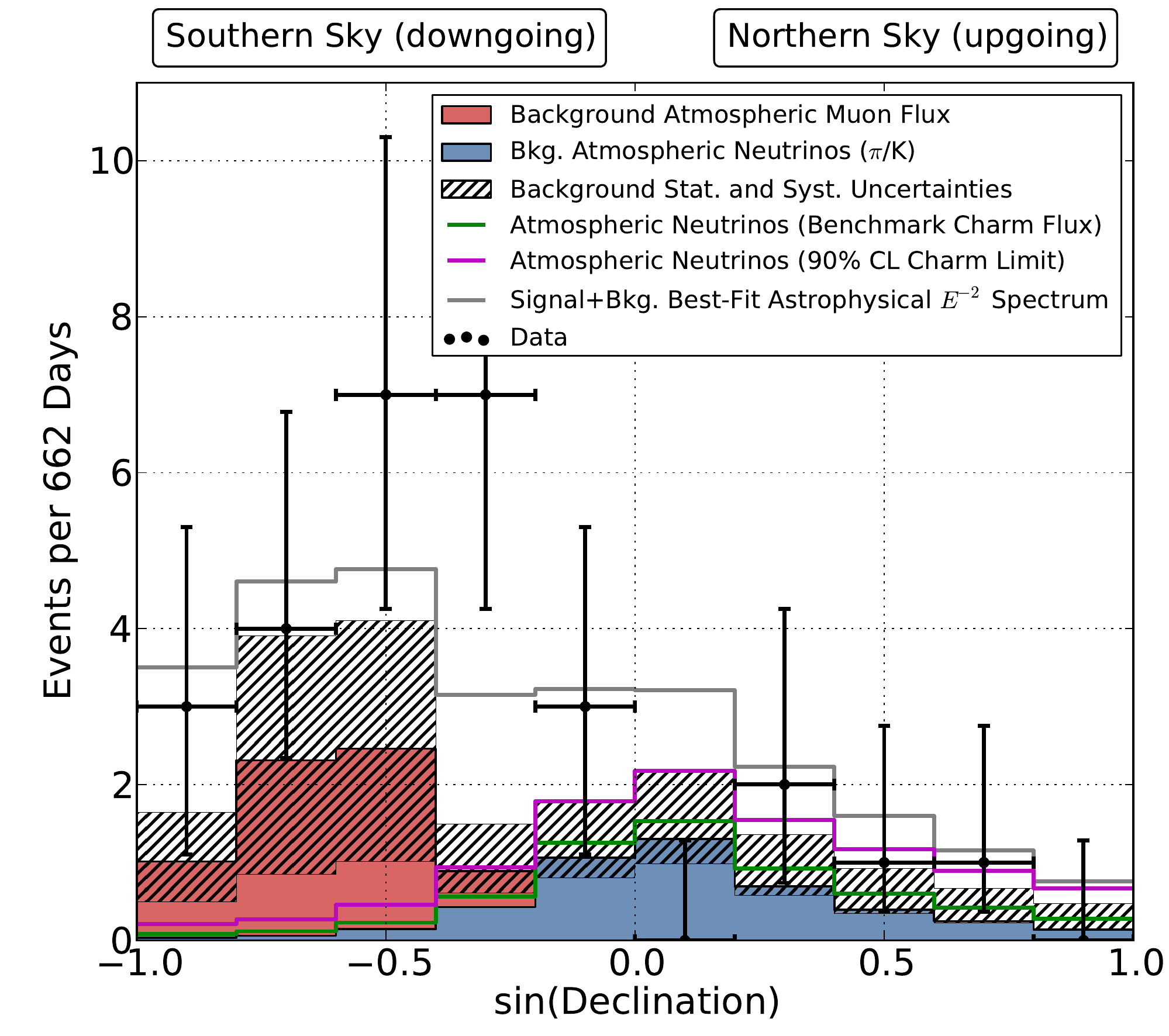}
\caption{%
Distributions of the deposited energies and declination angles of the observed events compared to model predictions.
Zenith angle entries for data (right) are the best-fit zenith position for each of the 28 events; a small number of events (Table~\ref{tab:events}) have zenith uncertainties larger than the bin widths in this figure.
Energies plotted (left) are reconstructed in-detector visible energies, which are lower limits on the neutrino energy.
Note that deposited energy spectra are always harder than the spectrum of the neutrinos that produced them due to the neutrino cross-section increasing with energy.
The expected rate of atmospheric neutrinos is shown in blue, with atmospheric muons in red.
The green line shows our benchmark atmospheric neutrino flux (see text), the magenta line the experimental 90\% bound.
Due to lack of statistics from data far above our cut threshold, the shape of the distributions from muons in this figure has been determined using Monte Carlo simulations with total rate normalized to the estimate obtained from our in-data control sample.
Combined statistical and systematic uncertainties on the sum of backgrounds are indicated with a hatched area.
The gray line shows the best-fit $E^{-2}$ astrophysical spectrum with a per-flavor normalization (1:1:1) of $E^2 \Phi_{\nu}(E) = 1.2 \cdot 10^{-8}\, \mathrm{GeV}\, \mathrm{cm}^{-2}\, \mathrm{s}^{-1}\, \mathrm{sr}^{-1}$.
}
\label{fig:spectrum}
\end{figure*}

Although there is some uncertainty in the expected atmospheric background rates, in particular for the contribution from charmed meson decays, the energy spectrum, zenith distribution, and shower to muon track ratio of the observed events strongly constrain the possibility that our events are entirely of atmospheric origin.
Almost all of the observed excess is in showers rather than muon tracks, ruling out an increase in penetrating muon background to the level required.
Atmospheric neutrinos are a poor fit to the data for a variety of reasons.
The observed events are much higher in energy, with a harder spectrum (Fig.~\ref{fig:spectrum}) than expected from an extrapolation of the well-measured $\pi/K$ atmospheric background at lower energies \cite{2011PhRvD..84h2001A,IC59DiffuseNOWPaper,dc_cascades}: nine had reconstructed deposited energies above 100 TeV, with two events above 1 PeV, relative to an expected background from $\pi/K$ atmospheric neutrinos of approximately 1 event above 100 TeV.
Raising the normalization of this flux both violates previous limits and, due to  $\nu_\mu$ bias in $\pi$ and $K$ decay, predicts too many muon tracks in our data ($2/3$ tracks vs. $1/4$ observed).

Another possibility is that the high-energy events result from charmed meson production in air showers \cite{EnbergPrompt,naumov}.
These produce higher energy events with equal parts $\nu_e$ and $\nu_\mu$, matching our observed muon track fraction reasonably well.
However, our event rates are substantially higher than even optimistic models \cite{naumov} and the energy spectrum from charm production is too soft to explain the data.
More importantly, increasing charm production to the level required to explain our observations violates existing experimental bounds \cite{IC59DiffuseNOWPaper}. 
As atmospheric neutrinos produced by any mechanism are made in cosmic ray air showers, downgoing atmospheric neutrinos from the southern sky will in general be accompanied into IceCube by muons produced in the same parent air shower.
These accompanying muons will trigger our muon veto, removing the majority of these events from the sample and biasing atmospheric neutrinos to the northern hemisphere.
The majority of our events, however, arrive from the south.
This places a strong model-independent constraint on any atmospheric neutrino production mechanism as an explanation for our data.

By comparison, a neutrino flux produced in extraterrestrial sources would, like our data, be heavily biased toward showers because neutrino oscillations over astronomical baselines tend to equalize neutrino flavors \cite{2009PhRvD..80k3006C,2008JHEP...02..005P}.
An equal-flavor $E^{-2}$ neutrino flux, for example, would be expected to produce only $1/5$ track events (see Materials and Methods).
The observed zenith distribution is also typical of such a flux:
as a result of absorption in the Earth above tens of TeV energy, most events (approximately $60\%$, depending on the energy spectrum) from even an isotropic high-energy extraterrestrial population would be expected to appear in the Southern Hemisphere.
Although the zenith distribution is well explained (Fig.~\ref{fig:spectrum}) by an isotropic flux, a slight southern excess remains, which could be explained either as a statistical fluctuation or by a source population that is either relatively small or unevenly distributed through the sky.

This discussion can be quantified by a global fit of the data to a combination of the $\pi/K$ atmospheric neutrino background, atmospheric neutrinos from charmed meson decays, and an isotropic equal-flavor extraterrestrial power-law flux.
With the normalizations of all components free to float, this model was fit to the two-dimensional deposited energy and zenith distribution of the data (Fig.~\ref{fig:energyzenith}) in the range 60 TeV $< E_\mathrm{dep}<$ 2 PeV, above the majority of the expected background (Fig. \ref{fig:spectrum}).
The data are well described in this energy range by an $E^{-2}$ neutrino spectrum with a per-flavor normalization of $E^2 \Phi(E) = (1.2 \pm 0.4) \cdot 10^{-8}\, \mathrm{GeV}\, \mathrm{cm}^{-2}\, \mathrm{s}^{-1}\, \mathrm{sr}^{-1}$.
Although it is difficult to substantively constrain the shape of the spectrum with our current limited statistics, a flux at this level would have been expected to generate an additional 3-6 events in the 2-10 PeV range; the lack of such events in the sample may indicate either a softer spectrum (the best fit is $E^{-2.2 \pm 0.4}$) or the presence of a break or cutoff at PeV energies.
When limited to only atmospheric neutrinos, the best fit to the data would require a charm flux 4.5 times larger than current experimental 90\% CL upper bounds \cite{IC59DiffuseNOWPaper} and even then is disfavored at $4 \sigma$ with respect to a fit allowing an extraterrestrial contribution.

\section*{Search for Neutrino Sources}
\label{sec:spatialclustering}

\begin{figure}
\includegraphics[width=0.99\linewidth]{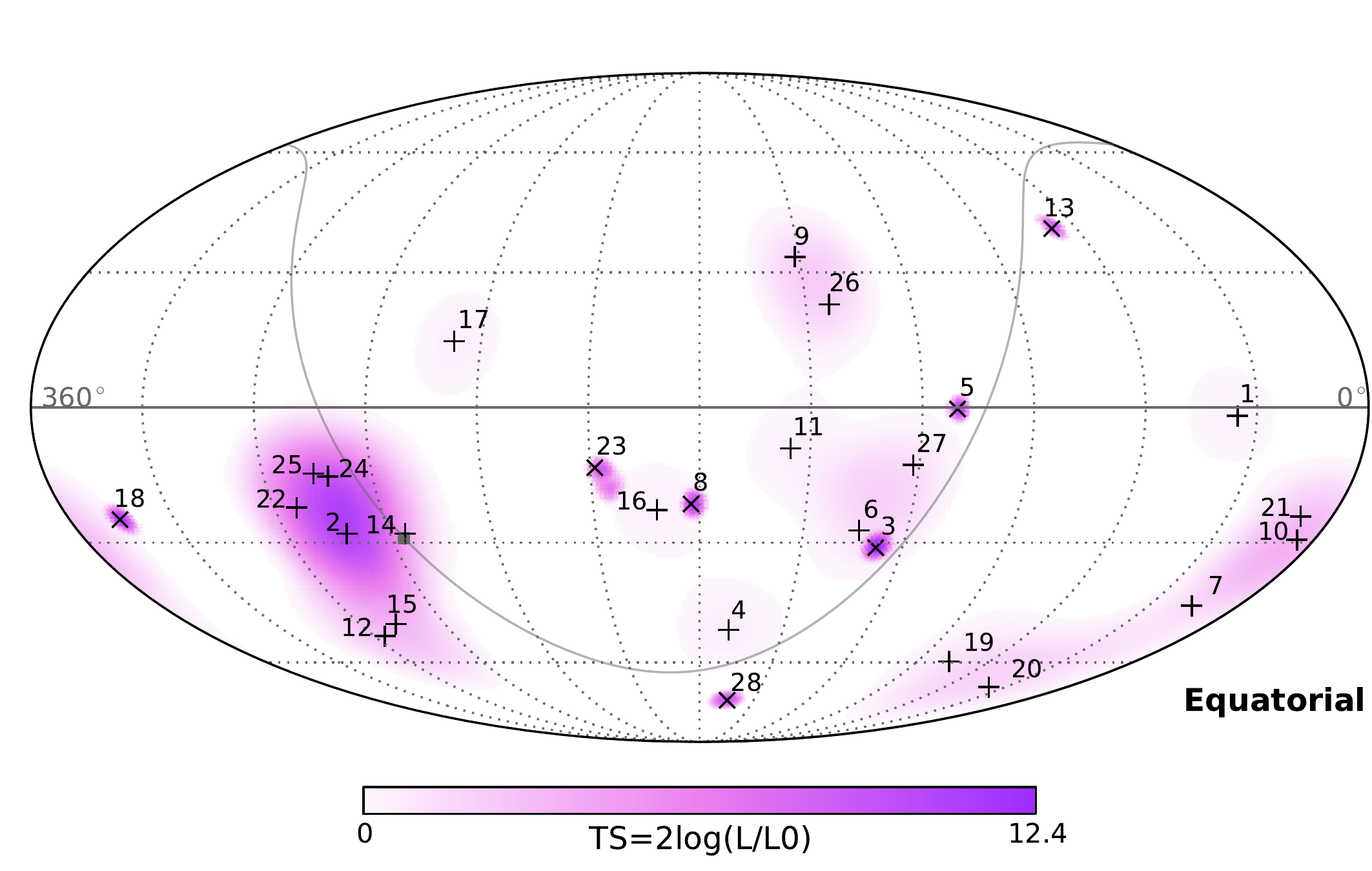}
\caption{%
Skymap in equatorial coordinates of the Test Statistic value (TS) from the maximum likelihood point-source analysis.
The most significant cluster consists of five events---all showers and including the second-highest energy event in the sample---with a final significance of 8\%.
This is not sufficient to identify any neutrino sources from the clustering study.
The galactic plane is shown as a gray line with the galactic center denoted as a filled gray square.
Best-fit locations of individual events (listed in Table~\ref{tab:events}) are indicated with vertical crosses ($+$) for showers and angled crosses ($\times$) for muon tracks.
}
\label{fig:skymap}
\end{figure}

In order to search for spatial clustering, indicating possible neutrino sources, we conducted a maximum likelihood point source analysis \cite{BraunLlh}.
At each point in the sky, we tested a point source hypothesis based on full-sky uncertainty maps for each event obtained from the reconstruction.
This yields a skymap of Test Statistic values (TS = $2\log(L/L_{0})$, where $L$ is the maximized likelihood and $L_{0}$ the likelihood under the null hypothesis), which reflects any excess concentration of events relative to a flat background distribution (Fig.~\ref{fig:skymap}).
To account for trials due to searching the whole sky, we estimate the significance of the highest TS observed by performing the same analysis on the data with the right ascension of the events randomized.
The final significance is then the fraction of these randomized maps that have a TS value anywhere in the sky as high or higher than that observed in data.
The chance probability calculated this way is independent of Monte Carlo simulation. 
Therefore, the significance obtained is against the hypothesis that all events in this sample are uniformly distributed in right ascension, rather than the significance of a cluster of events above predicted backgrounds.
Note that because muon tracks have much smaller angular uncertainties than showers, their presence can skew the highest TS values and overshadow clusters of shower events.
To correct for this effect, and because muon events are more likely to be atmospheric background, every clustering analysis described here was repeated twice: once with the full 28 events and once with only the 21 shower events.

When using all events, the likelihood map reveals no significant clustering compared to randomized maps.
For the shower events, the coordinates with the highest TS are at RA=$281^{\circ}$, dec=$-23^{\circ}$ (galactic longitude $l=+12^{\circ}$, latitude $b=-9^{\circ}$).
Five events, including the second highest energy event in the sample, contribute to the main part of the excess with two others nearby.
The fraction of randomized data sets which yield a similar or higher TS at this exact spot is $0.2\%$.
(At the exact location of the Galactic Center, the fraction is $5.4\%$.)
The final significance, estimated as the fraction of randomized maps with a similar or higher TS anywhere in the sky, is $8\%$.
This degree of clustering may be compatible with a source or sources in the galactic center region but the poor angular resolution for showers and wide distribution of the events do not allow the identification of any sources at this time.

Two other spatial clustering analyses were defined a priori.
We performed a galactic plane correlation study using the full directional reconstruction uncertainty for each event to define the degree of overlap with the plane. 
The plane width was chosen to be $\pm2.5^{\circ}$ following TeV gamma-ray observations \cite{milagroGP}.
A multi-cluster search using the sum of log-likelihood values at every local maximum in the likelihood map was also conducted.
Neither of these analyses yielded significant results.

In addition to clustering of events in space, we performed two tests for clustering of events in time that calculate significances by comparing the actual arrival times to event times drawn from a random uniform distribution throughout the live time.
Because many sources \cite{wb97,2005ApJ...633.1013I,2012ApJ...745...45A} are expected to produce neutrinos in bursts, identification of such a time cluster could allow association with a source without reference to the limited angular resolution of the majority of the observed neutrinos.
When using all events, no significant time cluster was observed.
Furthermore, each spatial cluster in Fig.~\ref{fig:skymap} containing more than one event was tested individually for evidence of time clustering.
Of the eight regions tested, the most significant was a pair that includes the highest energy shower in the sample, but was still compatible with random fluctuations.
The five shower events of the densest cluster show no significant overall time clustering.



\section*{Materials and Methods}

\subsection*{Event Selection}
\label{sec:event_selection}

Backgrounds for cosmic neutrino searches arise entirely from interactions of cosmic rays in the Earth's atmosphere.
These produce secondary muons that penetrate into underground neutrino detectors from above as well as atmospheric neutrinos that reach the detector from all directions due to the low neutrino cross-section which allows them to penetrate the Earth from the opposite hemisphere.
These particles are produced in the decays of secondary $\pi$ and $K$ mesons; at high energies a flux from the prompt decay of charmed mesons \cite{charm} has been anticipated although not yet observed.
Cosmic ray muons are the dominant background in IceCube due to their high rate of 3 kHz.
These can be removed from the sample either by using only upgoing events, by limiting searches to events at very high energies (above about 1 PeV) \cite{2009PhRvL.103v1102A,ic40_ehe}, or, as here, by requiring an observation of the neutrino interaction vertex using the detector boundary to detect and veto entering muon tracks.

Neutrino candidates were selected by finding events that originated within the detector interior.
Included were those events that produced their first light within the fiducial volume (Fig.~\ref{fig:icecube_veto}) and were of sufficiently high energy such that an entering muon track would have been reliably identified if present.
In particular, we required that each event have fewer than three of its first 250 observed photoelectrons (p.e.) detected in the veto region.
In addition, we required that the event produce at least 6000 p.e. overall to ensure that statistical fluctuations in the light yield were low enough for entering muons to reliably produce light in the veto region.
This event selection rejects 99.999\% of the muon background above 6000 p.e. (Fig.~\ref{fig:taggedvetorates}) while retaining nearly all neutrino events interacting within the fiducial volume at energies above a few hundred TeV.
This selection is largely independent of neutrino flavor, event topology, or arrival direction.
It also removes 70\% of atmospheric neutrinos \cite{atmonu_veto} in the Southern Hemisphere, where atmospheric neutrinos are usually accompanied into the detector by muons produced in the same parent air shower.
To prevent confirmation bias, we conducted a blind analysis designed on a subsample of 10\% of the full dataset.

\subsection*{Event Reconstruction}
\label{sec:event_reco}

Neutrino interactions in IceCube have two primary topologies: showers and muon tracks.
Showers are created by secondary leptons and hadronic fragmentation in $\nu_e$ and $\nu_\tau$ charged-current interactions and by neutral-current interactions of neutrinos of all flavors.
At the relevant energies ($\gtrsim$ 50 TeV), showers, including tracks left by $\tau$ leptons, have a length of roughly 10 meters in ice and are, to a good approximation, point sources of light \cite{cascade_light}.
Secondary muon tracks are created primarily in $\nu_\mu$ charged-current interactions along with a hadronic shower at the neutrino interaction vertex, and have a typical range on the order of kilometers, larger than the dimensions of the detector.
Note that, for a flux consisting of a mixture of flavors, this implies that showers will be the dominant topology since $\nu_\mu$ CC represents only a small fraction of the total event rate:
for an equally mixed $E^{-2}$ spectrum, approximately 80\% of the observed events would appear as showers.

Although the distribution of hit PMTs in the detector is approximately spherical for shower events, the detailed timing patterns of the photons in the individual PMTs retain the memory of the direction of the primary lepton.
Comparison of these distributions with expectations from simulated showers yields a typical median angular resolution of $10^\circ$-$15^\circ$.
Resolution on deposited energy, from the recorded waveform amplitudes, is typically 10-15\%.
In events with a muon track, the extension of the track in the detector provides a much tighter constraint on direction than the shapes of the waveforms alone, improving angular resolution greatly to better than $1^\circ$ \cite{moonpaper}.
Energy reconstruction only yields a lower limit on neutrino energy as a result of the energy removed from the detector by escaping muons and neutrinos.
All quoted directional and energy reconstruction uncertainties are dominated by a systematic component arising from uncertainties in the optical properties of the ice \cite{spice} and the optical sensitivity of the PMTs \cite{PMT_paper}.

\subsection*{Atmospheric Muon Background}
\label{sec:background_estimation_muon}

\begin{figure}
\includegraphics[width=\linewidth]{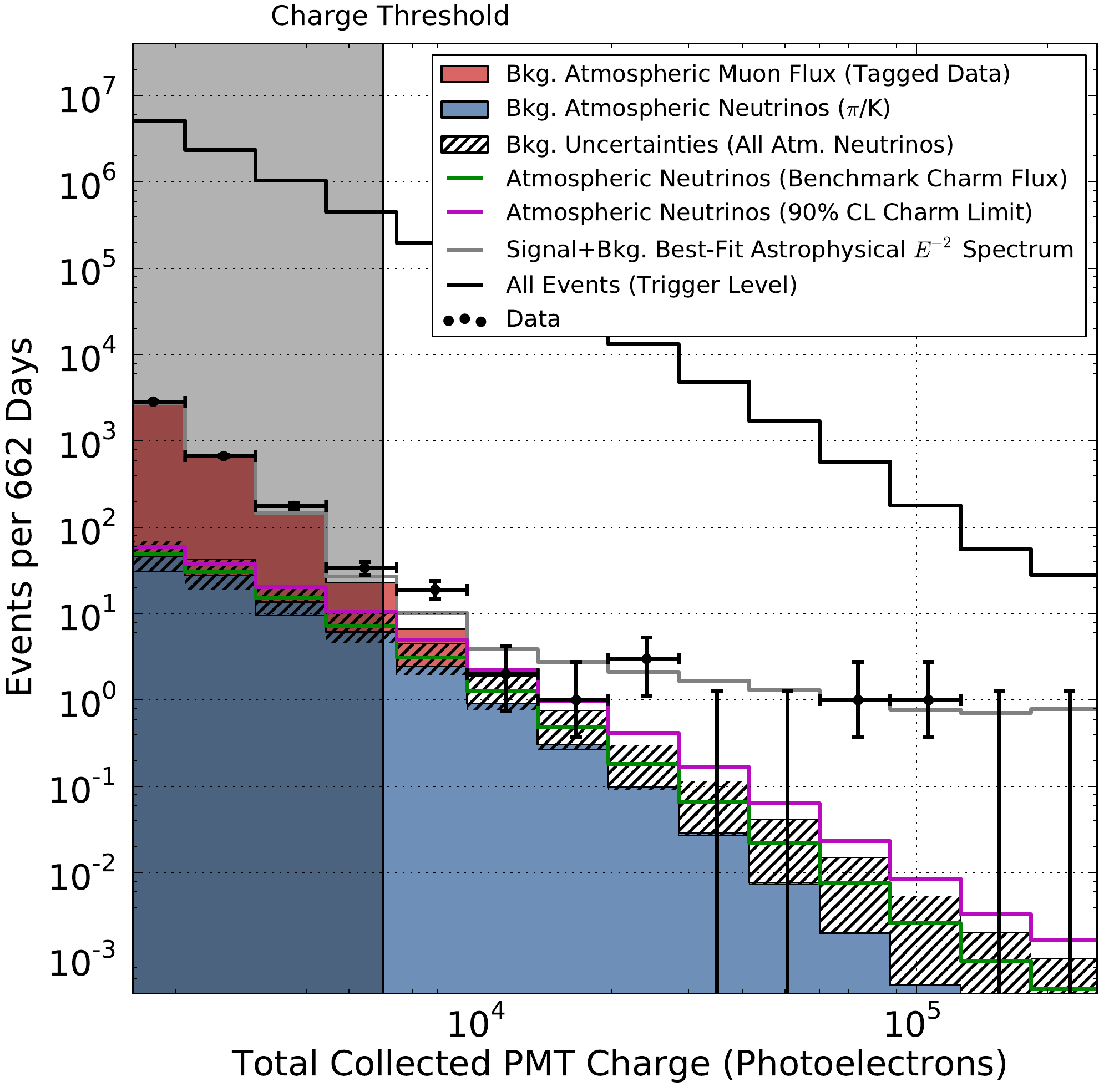}
\caption{%
Distribution of deposited PMT charges ($Q_{tot}$).
Muons at higher total charges are less likely to pass the veto layer undetected, causing the muon background (red, estimated from data) to fall faster than the overall trigger rate (uppermost line).
The data events in the unshaded region, at $Q_{tot} > 6000$, are the events reported in this work, with error bars indicating 68\% Feldman-Cousins intervals.
The best-fit $E^{-2}$ astrophysical spectrum (gray line) and atmospheric neutrino flux (blue) have been determined using Monte Carlo simulations, with the hatched region showing current experimental uncertainties on the atmospheric neutrino background.
The largest of these uncertainties is neutrinos from charmed meson decays, a flux which has yet to be observed and is thus not included in the blue region; the hatched region includes the best experimental $1 \sigma$ upper limit \cite{IC59DiffuseNOWPaper}.
For scale, two specific charm levels are also shown: a benchmark theoretical model \cite{EnbergPrompt} (green line) and the experimental 90\% CL upper bound \cite{IC59DiffuseNOWPaper} (magenta line).
}
\label{fig:taggedvetorates}
\end{figure}

\begin{figure}
\includegraphics[width=\linewidth]{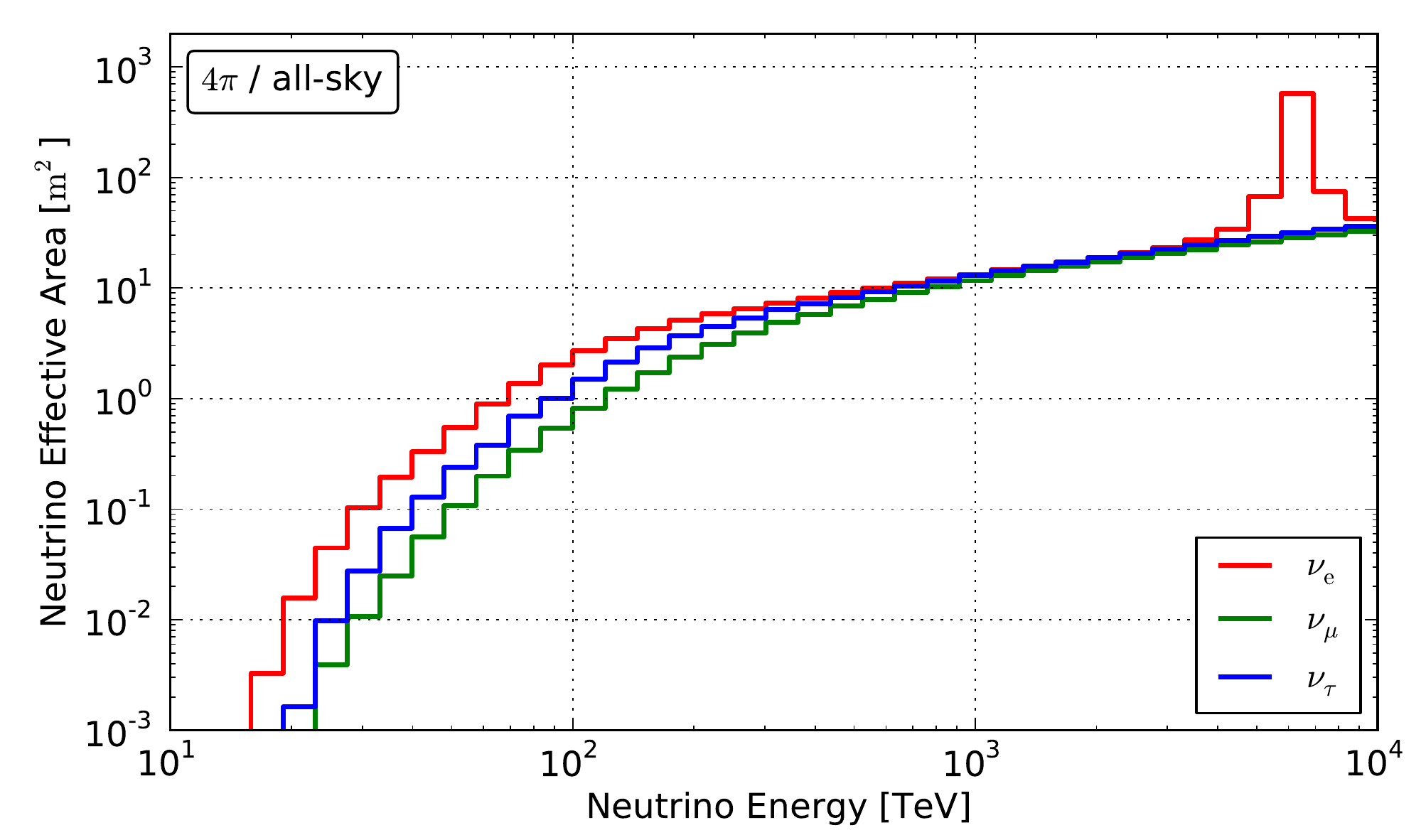}
\includegraphics[width=\linewidth]{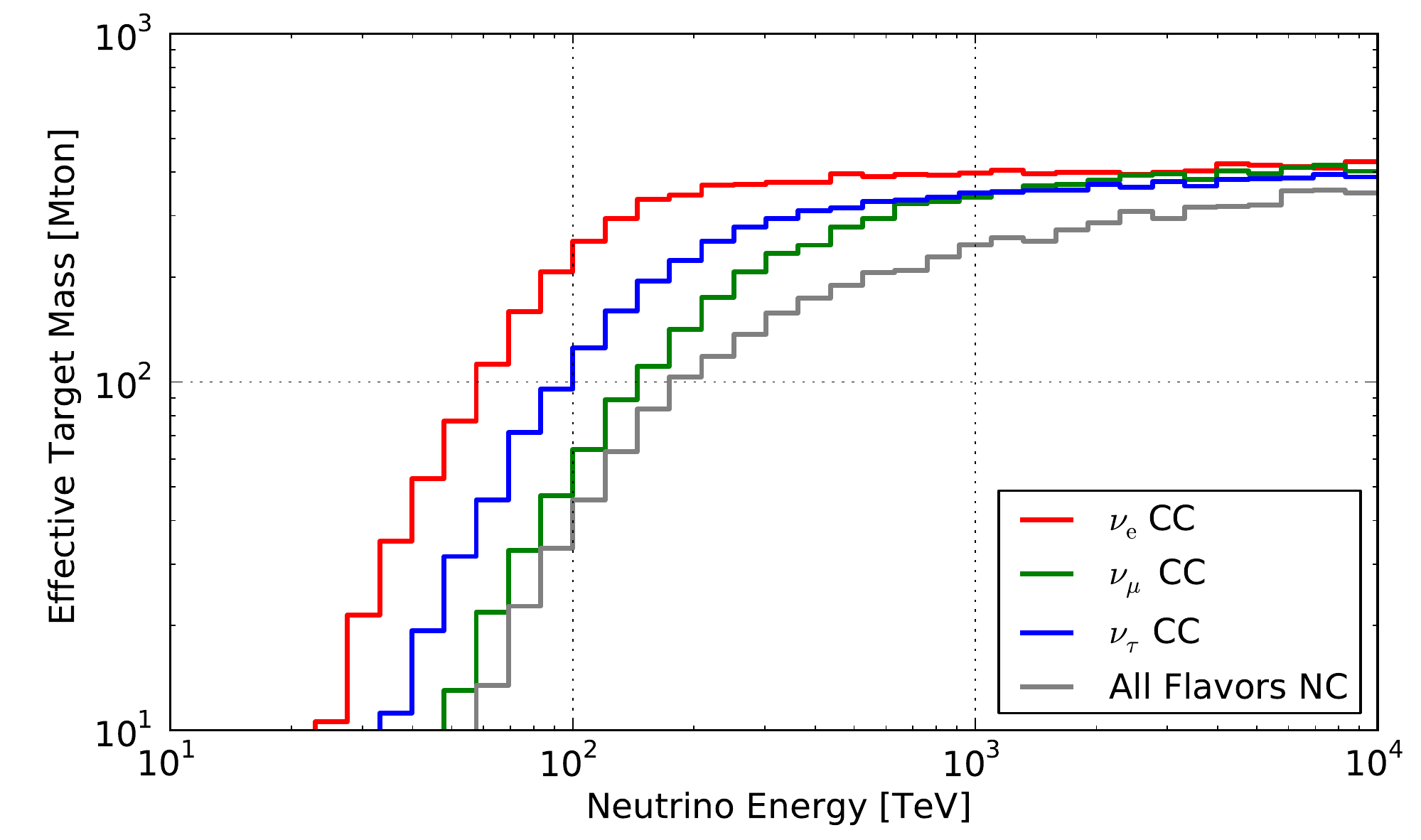}
\caption{%
Neutrino effective area and volume.
Event rates can be obtained by multiplying the effective areas by $4\pi$, by the sum of $\nu$ and $\bar{\nu}$ fluxes, and by the livetime of 662 days.
Top:
Neutrino effective areas for each flavor assuming an equal flux of neutrinos and antineutrinos and averaged over all arrival angles.
At 6.3 PeV, resonant $W$ production on atomic electrons increases sensitivity to $\bar \nu_e$.
The effective area includes effects from attenuation of neutrinos in the Earth \cite{css_crosssections}, relevant at energies above 100 TeV.
Bottom:
Effective target mass as a function of energy.
The deposited energy threshold in this search causes some flavor bias at low energies due to missing energy in escaping particles from $\nu_\mu$ and $\nu_\tau$ charged-current events.
For $\nu_e$ charged-current events, where all the neutrino energy is visible in the detector, full efficiency is reached above 100 TeV.
}
\label{fig:effective_area}
\end{figure}

Remaining atmospheric muon background comes from tracks that produce too little light at the edge of the detector to be vetoed and instead emit their first detected photons in the interior volume, mimicking a starting neutrino.
These events usually produce an observable muon track in the detector like that from a $\nu_\mu$ charged-current event.
Much more rarely, catastrophic energy loss processes such as muon bremsstrahlung can create a shower-like signal, especially in the corners of the detector where the exiting muon track may not be observed.

The veto passing rate for throughgoing muons, and therefore the total muon background in the analysis, can be evaluated directly from the data by implementing a two-layer anticoincidence detector.
Entering events can be tagged with high efficiency using the outer layer of IceCube; the rate of these tagged events that pass the next veto layer can be used as a control sample to evaluate the rate at which muons are detected by a single detector layer as a function of observed light yield.
This per-layer probability can be used to estimate the final background rate after application of a geometrical correction factor of approximately a factor of two for the larger size of the analysis fiducial volume compared to the deep interior fiducial volume (after two veto layers).
The resulting predicted veto passing rate agrees well with data at low energies where we expect the event rate to be background dominated (Fig.~\ref{fig:taggedvetorates}).
In our signal region above 6000 p.e., we observed three tagged events passing the inner veto and so predict $6.0 \pm 3.4$ veto-penetrating muon events in the two-year data set.

\subsection*{Atmospheric Neutrinos}
\label{sec:background_estimation_neutrino}

Atmospheric neutrino backgrounds, including an as-yet unobserved component from charmed meson decays, were estimated based on a parametrization of the atmospheric neutrino flux \cite{Honda2006,GaisserImprovedCRSpectrum} consistent with previous IceCube measurements of northern-hemisphere muon neutrinos \cite{IC59DiffuseNOWPaper}.
We have also included a suppression of the atmospheric neutrino background from the Southern Hemisphere resulting from the fact that accompanying high-energy muons produced in the same air shower can trigger our muon veto if they penetrate to the depth of the detector.
Here we have extended previous analytic calculations \cite{atmonu_veto} of this suppression factor using the CORSIKA \cite{CORSIKA} air-shower simulation to determine the fraction of atmospheric neutrinos accompanied at depth by muons above 10 TeV, at which they will be reliably detected by our muon veto.
This factor is a strong function of neutrino energy and angle, with the strongest suppression expected at high energies and most downward angles.
The suppression factor, bounded above at 90\% to cover uncertainties in hadronic interaction models, was then folded with the northern-hemisphere spectrum to predict the southern-hemisphere event rate.

This produces an estimate of the atmospheric neutrino background of $4.6^{+3.7}_{-1.2}$ events in the 662 day livetime.
These events would be concentrated near the energy threshold of the analysis due to the steeply falling atmospheric neutrino spectrum.
Uncertainties in the atmospheric neutrino background are dominated by the flux from charmed meson decays, which is too small to have been observed thus far and is currently bounded above experimentally by a $1\sigma$ upper limit of 3.4 events~\cite{IC59DiffuseNOWPaper}.
The spectrum and composition of cosmic rays and models of hadronic interactions contribute a rate uncertainty at the relevant energies of $_{-20\%}^{+30\%}$, which dominates the uncertainties in the $\pi/K$ component of the spectrum \cite{2012PhRvD..86k4024F}.
The measured $5\%$ uncertainty in the electromagnetic energy scale and detector linearity contributes a proportional $\pm 15\%$ uncertainty to the atmospheric background rates.
Given the charge threshold, uncertainty in the light yield of hadronic showers, which is less well constrained, can affect the estimated background neutrino rate. However, the light yield for a hadronic shower is smaller than the well known light yield for an electromagnetic shower at the same energy, limiting any resulting increase in the background rate to no larger than $30\%$.

\subsection*{Blind Calculation of Significance}

We evaluated the significance of the excess over atmospheric backgrounds based on both the total rate and properties of the observed events.
From each event, the total deposited PMT charge, reconstructed energy, and direction were used to compute tail probabilities relative to the atmospheric muon and neutrino backgrounds.
Overall significance was computed using the product of the per-event probabilities as a test statistic.

The muon background probability was computed as the fraction of the expected background with deposited charge greater than observed.
Above the highest charge event in the control sample, we set an upper limit on the passing rate by assuming a constant veto efficiency.
Similarly, the likelihood ratio between an isotropic $E^{-2}$ astrophysical flux and the expected atmospheric neutrino background in declination and deposited energy was calculated for each event after folding with the observed reconstruction uncertainties, and the probability for an atmospheric neutrino event to have a larger value than observed was computed.
Because our control sample of background muon events has limited statistics, we cannot produce a detailed map of the energies and angles of the penetrating muon background.
For this reason, the muon and neutrino background probabilities were combined by taking the maximum of the two as the statistic for each event, which will somewhat underestimate the significance of any excess.

\begin{acknowledgments}
We acknowledge support from the following agencies:
U.S. National Science Foundation-Office of Polar Programs,
U.S. National Science Foundation-Physics Division,
University of Wisconsin Alumni Research Foundation,
the Grid Laboratory Of Wisconsin (GLOW) grid infrastructure at the University of Wisconsin - Madison, the Open Science Grid (OSG) grid infrastructure;
U.S. Department of Energy, and National Energy Research Scientific Computing Center,
the Louisiana Optical Network Initiative (LONI) grid computing resources;
Natural Sciences and Engineering Research Council of Canada,
WestGrid and Compute/Calcul Canada;
Swedish Research Council,
Swedish Polar Research Secretariat,
Swedish National Infrastructure for Computing (SNIC),
and Knut and Alice Wallenberg Foundation, Sweden;
German Ministry for Education and Research (BMBF),
Deutsche Forschungsgemeinschaft (DFG),
Helmholtz Alliance for Astroparticle Physics (HAP),
Research Department of Plasmas with Complex Interactions (Bochum), Germany;
Fund for Scientific Research (FNRS-FWO),
FWO Odysseus programme,
Flanders Institute to encourage scientific and technological research in industry (IWT),
Belgian Federal Science Policy Office (Belspo);
University of Oxford, United Kingdom;
Marsden Fund, New Zealand;
Australian Research Council;
Japan Society for Promotion of Science (JSPS);
the Swiss National Science Foundation (SNSF), Switzerland;
National Research Foundation of Korea (NRF)
\end{acknowledgments}

Additional data and resources are available in the supporting online material (\url{http://www.sciencemag.org/content/342/6161/1242856/suppl/DC1}).
These include displays of the neutrino candidate events and list precise arrival times, as well as machine-readable tabular neutrino effective areas (Fig.~\ref{fig:effective_area}).
IceCube data are archived at \url{http://www.icecube.wisc.edu/science/data}.

\ifx \includeeventviews\undefined
\else
\onecolumngrid
\appendix
\newpage
\section*{Event 1}

\includegraphics[width=0.8\linewidth]{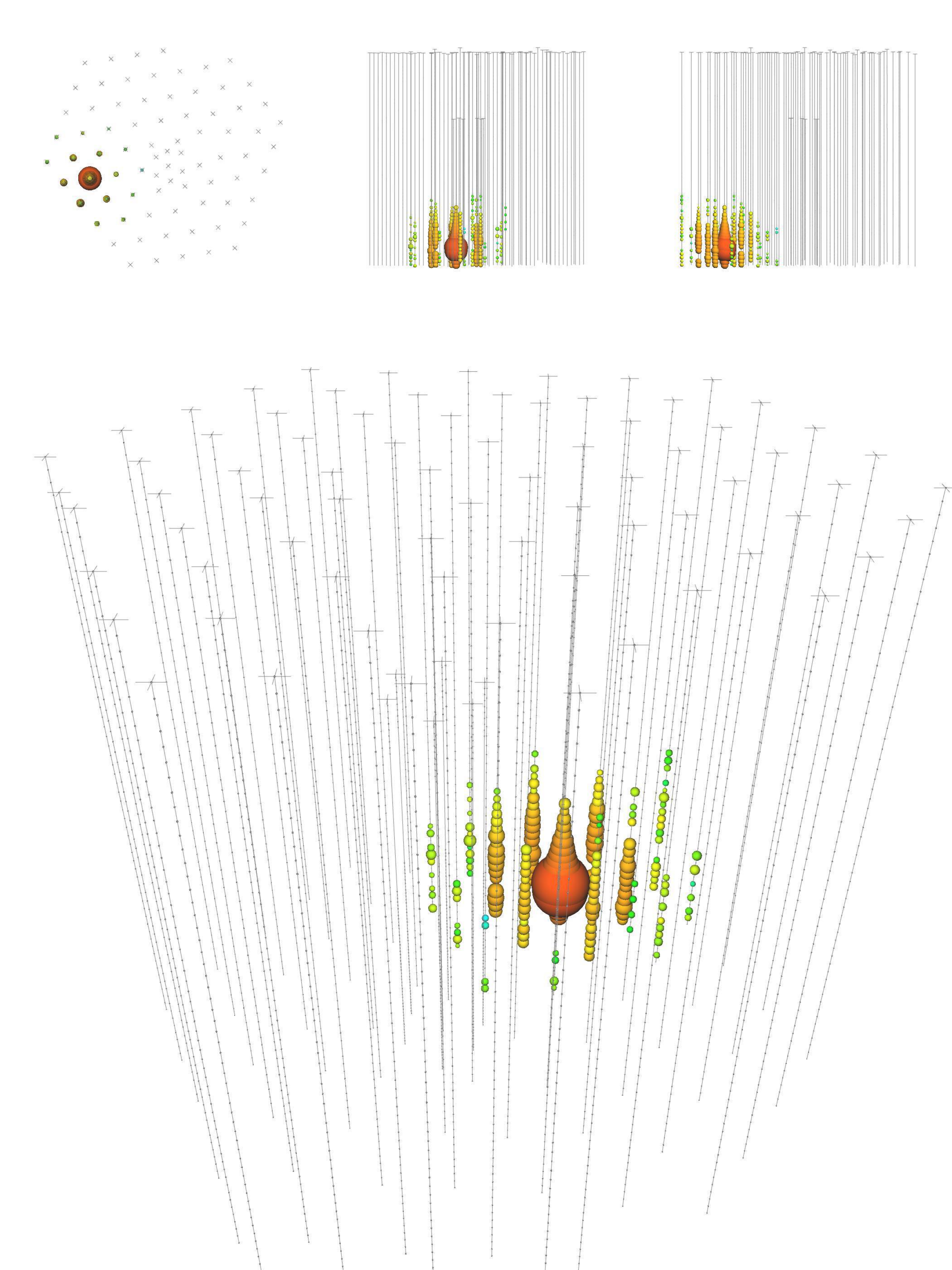}\\
\\
\includegraphics[width=0.8\linewidth]{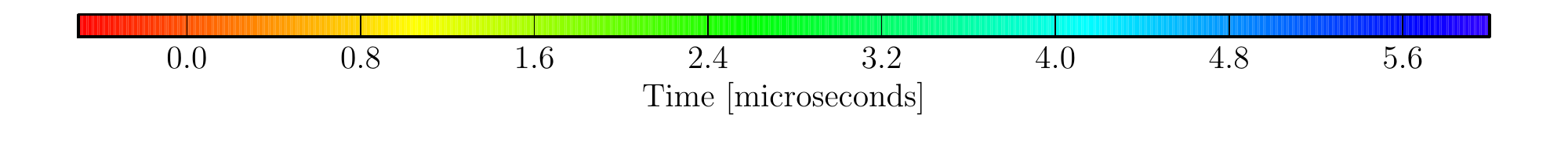}\vspace{0.2in}\\
\begin{tabular}{c|c|c|c|c|c}
Deposited Energy (TeV) & Time (MJD) & Declination (deg.) & RA (deg.) & Med. Ang. Resolution (deg.) & Topology\\
\hline
$47.6 \,^{+6.5}_{-5.4}$ & 55351.3222110 & $-1.8$ & $35.2$ & $16.3$ & Shower
\end{tabular}
\newpage
\section*{Event 2}

\includegraphics[width=0.8\linewidth]{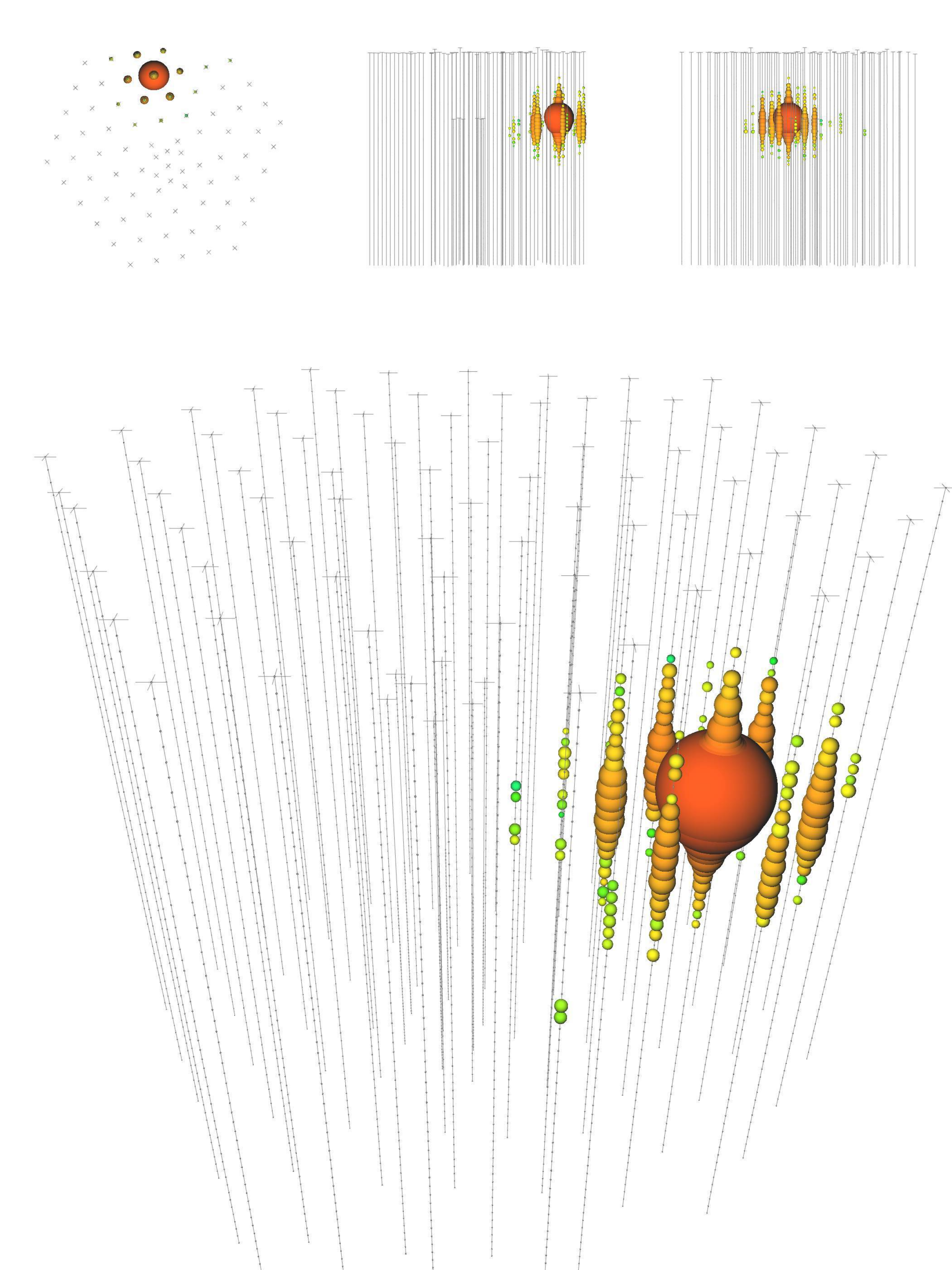}\\
\\
\includegraphics[width=0.8\linewidth]{color_scale.pdf}\vspace{0.2in}\\
\begin{tabular}{c|c|c|c|c|c}
Deposited Energy (TeV) & Time (MJD) & Declination (deg.) & RA (deg.) & Med. Ang. Resolution (deg.) & Topology\\
\hline
$117.0 \,^{+15.4}_{-14.6}$ & 55351.4659612 & $-28.0$ & $282.6$ & $25.4$ & Shower
\end{tabular}
\newpage
\section*{Event 3}

\includegraphics[width=0.8\linewidth]{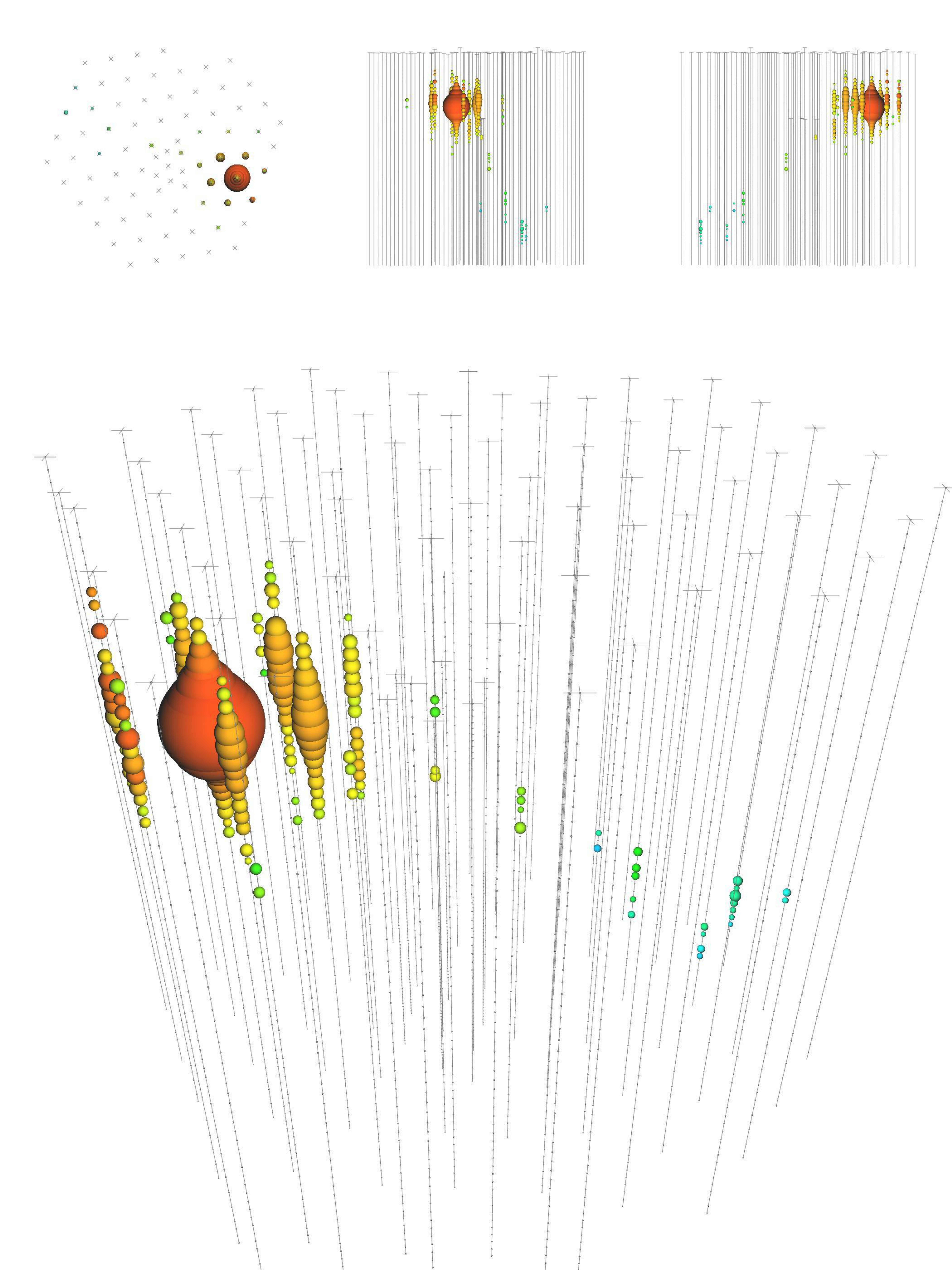}\\
\\
\includegraphics[width=0.8\linewidth]{color_scale.pdf}\vspace{0.2in}\\
\begin{tabular}{c|c|c|c|c|c}
Deposited Energy (TeV) & Time (MJD) & Declination (deg.) & RA (deg.) & Med. Ang. Resolution (deg.) & Topology\\
\hline
$78.7 \,^{+10.8}_{-8.7}$ & 55451.0707415 & $-31.2$ & $127.9$ & $\lesssim 1.4$ & Track
\end{tabular}
\newpage
\section*{Event 4}

\includegraphics[width=0.8\linewidth]{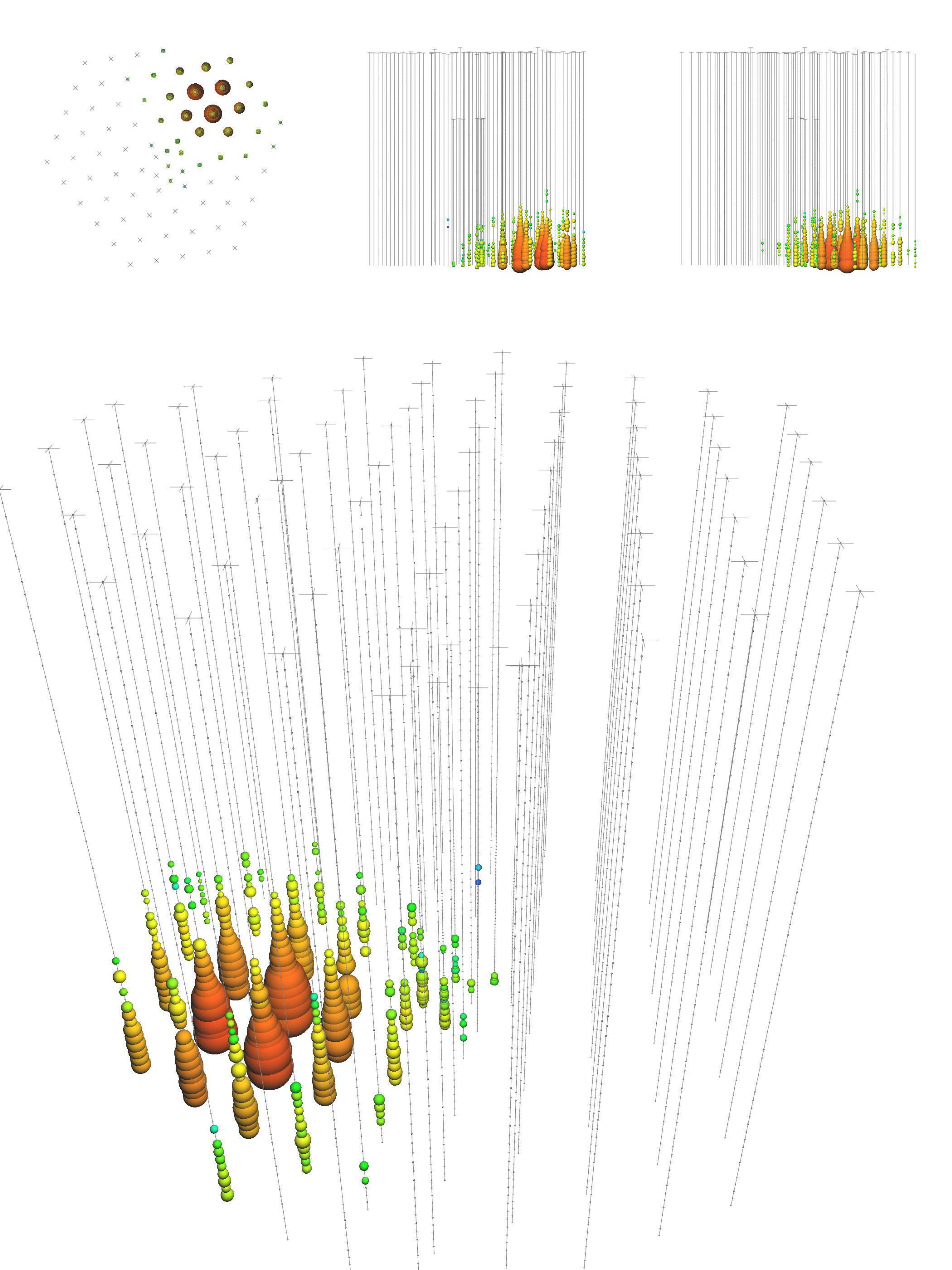}\\
\\
\includegraphics[width=0.8\linewidth]{color_scale.pdf}\vspace{0.2in}\\
\begin{tabular}{c|c|c|c|c|c}
Deposited Energy (TeV) & Time (MJD) & Declination (deg.) & RA (deg.) & Med. Ang. Resolution (deg.) & Topology\\
\hline
$165.4 \,^{+19.8}_{-14.9}$ & 55477.3930911 & $-51.2$ & $169.5$ & $7.1$ & Shower
\end{tabular}
\newpage
\section*{Event 5}

\includegraphics[width=0.8\linewidth]{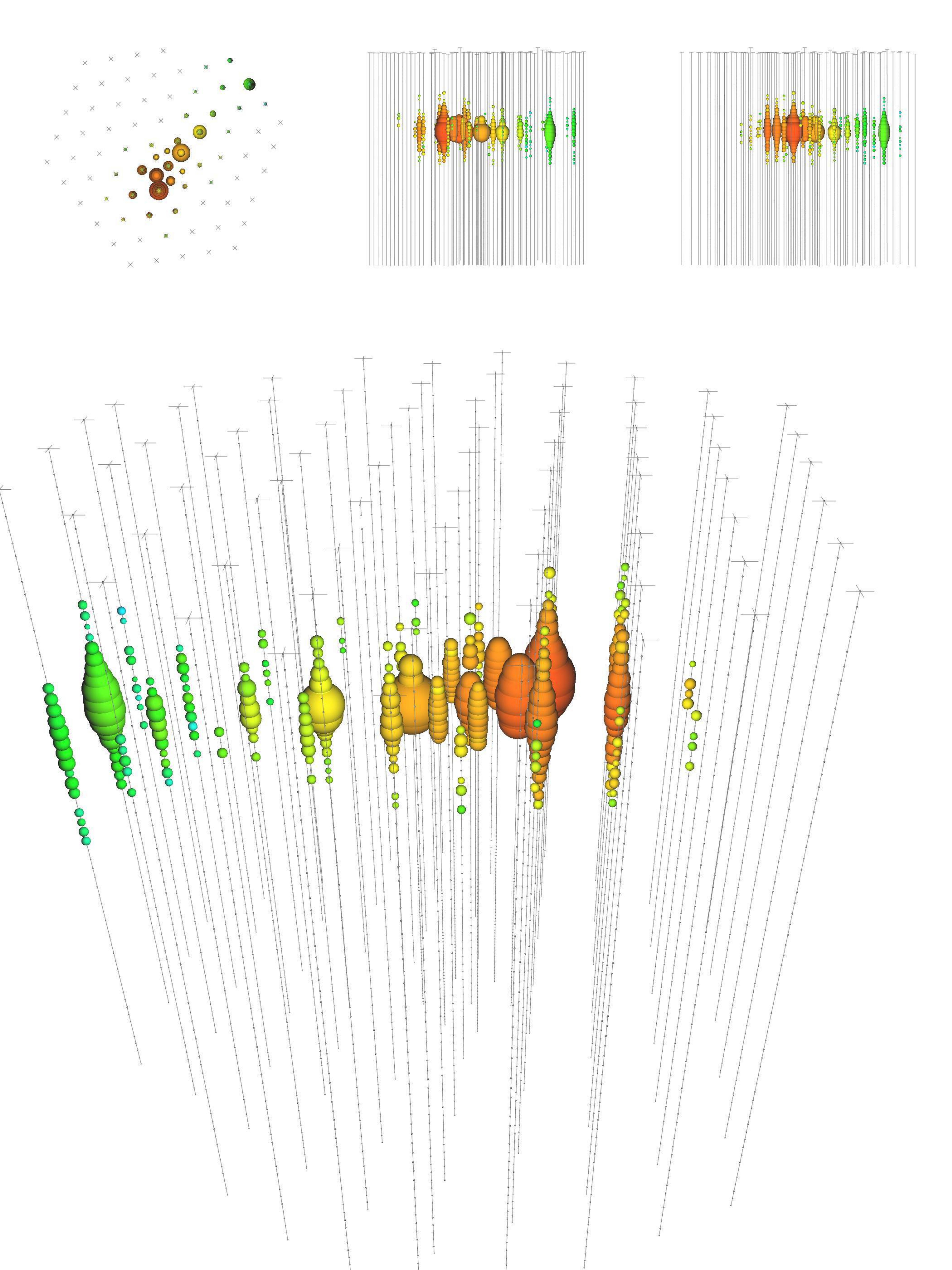}\\
\\
\includegraphics[width=0.8\linewidth]{color_scale.pdf}\vspace{0.2in}\\
\begin{tabular}{c|c|c|c|c|c}
Deposited Energy (TeV) & Time (MJD) & Declination (deg.) & RA (deg.) & Med. Ang. Resolution (deg.) & Topology\\
\hline
$71.4 \,^{+9.0}_{-9.0}$ & 55512.5516214 & $-0.4$ & $110.6$ & $\lesssim 1.2$ & Track
\end{tabular}
\newpage
\section*{Event 6}

\includegraphics[width=0.8\linewidth]{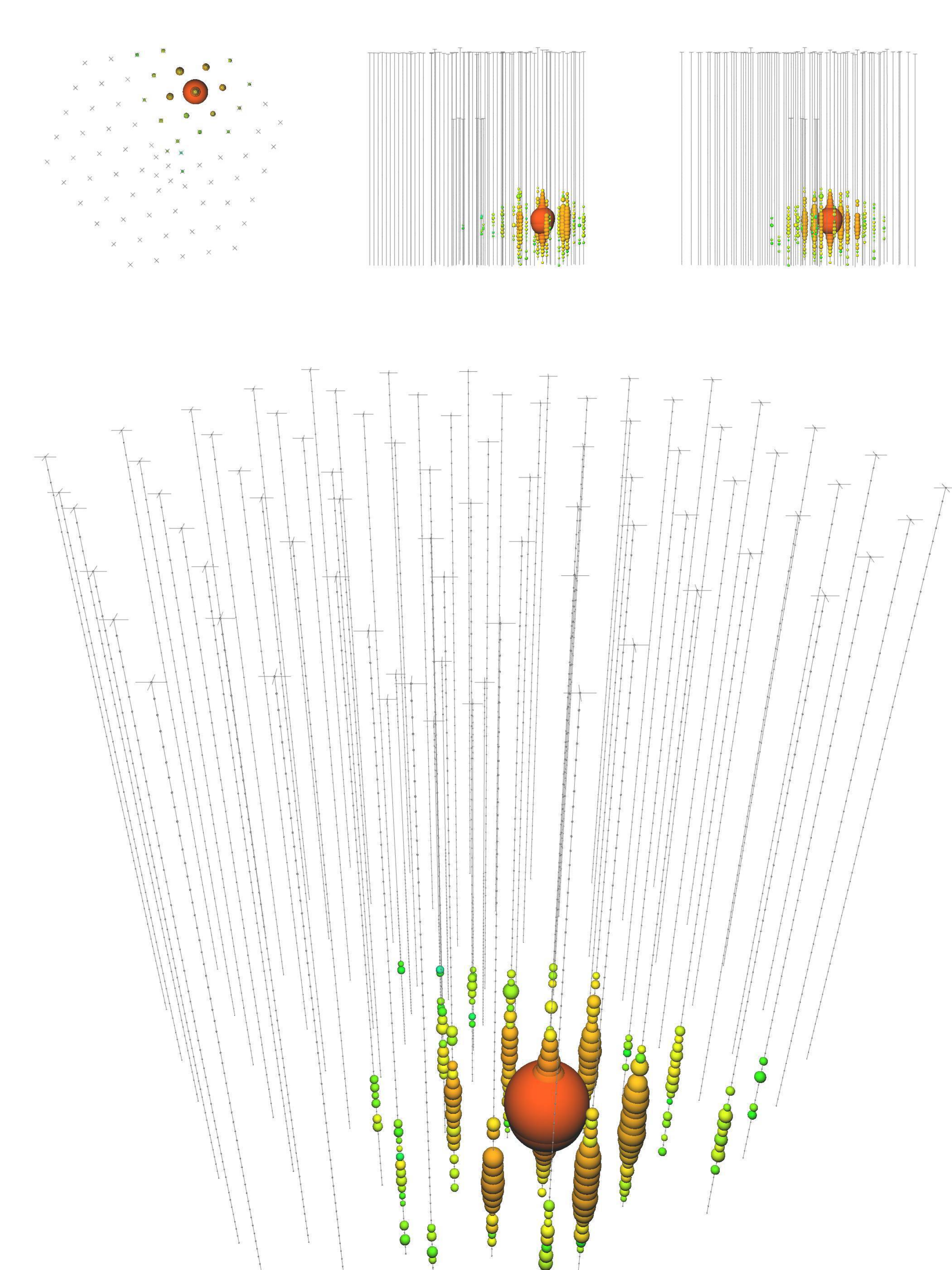}\\
\\
\includegraphics[width=0.8\linewidth]{color_scale.pdf}\vspace{0.2in}\\
\begin{tabular}{c|c|c|c|c|c}
Deposited Energy (TeV) & Time (MJD) & Declination (deg.) & RA (deg.) & Med. Ang. Resolution (deg.) & Topology\\
\hline
$28.4 \,^{+2.7}_{-2.5}$ & 55567.6388084 & $-27.2$ & $133.9$ & $9.8$ & Shower
\end{tabular}
\newpage
\section*{Event 7}

\includegraphics[width=0.8\linewidth]{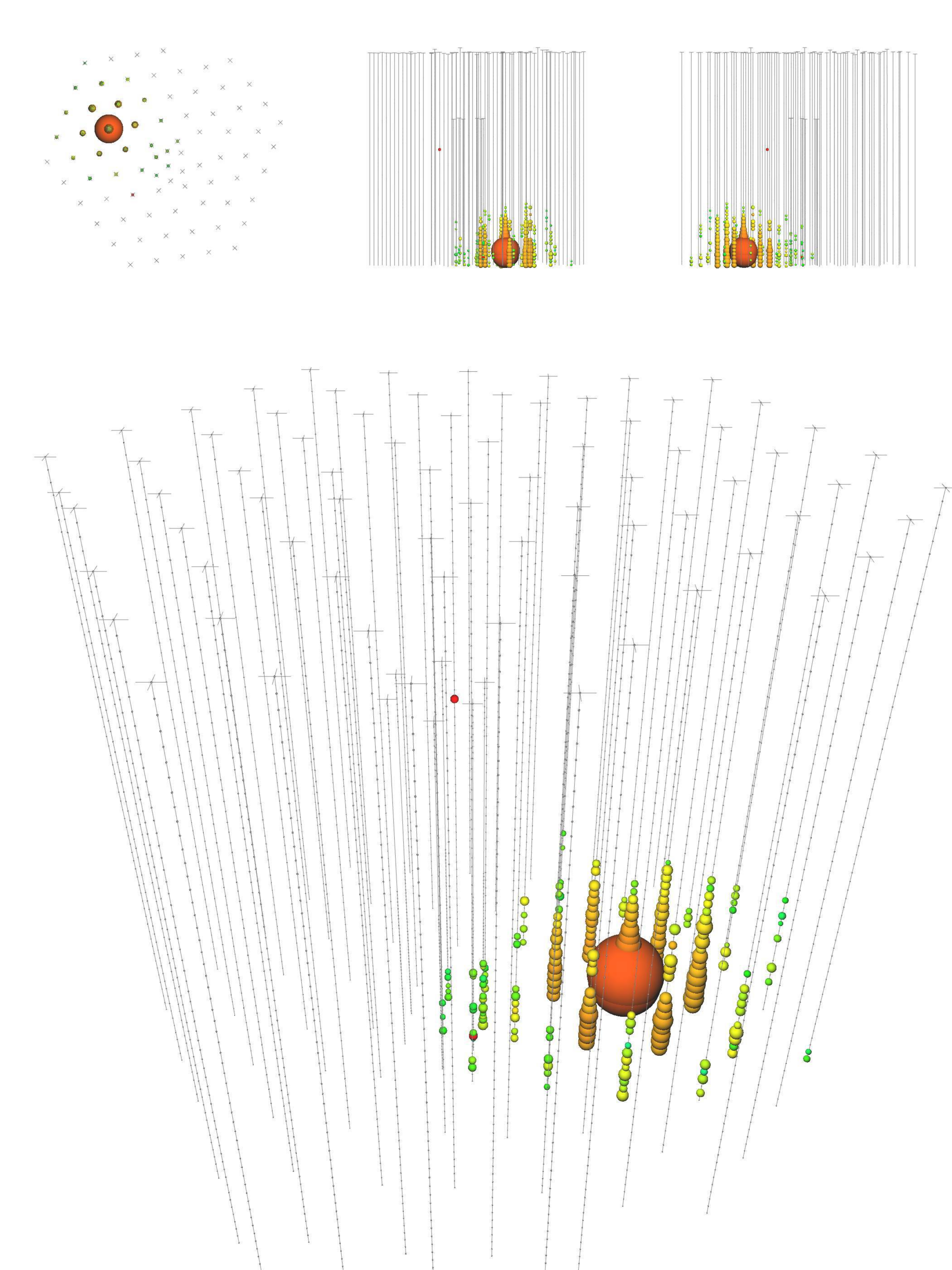}\\
\\
\includegraphics[width=0.8\linewidth]{color_scale.pdf}\vspace{0.2in}\\
\begin{tabular}{c|c|c|c|c|c}
Deposited Energy (TeV) & Time (MJD) & Declination (deg.) & RA (deg.) & Med. Ang. Resolution (deg.) & Topology\\
\hline
$34.3 \,^{+3.5}_{-4.3}$ & 55571.2585307 & $-45.1$ & $15.6$ & $24.1$ & Shower
\end{tabular}
\newpage
\section*{Event 8}

\includegraphics[width=0.8\linewidth]{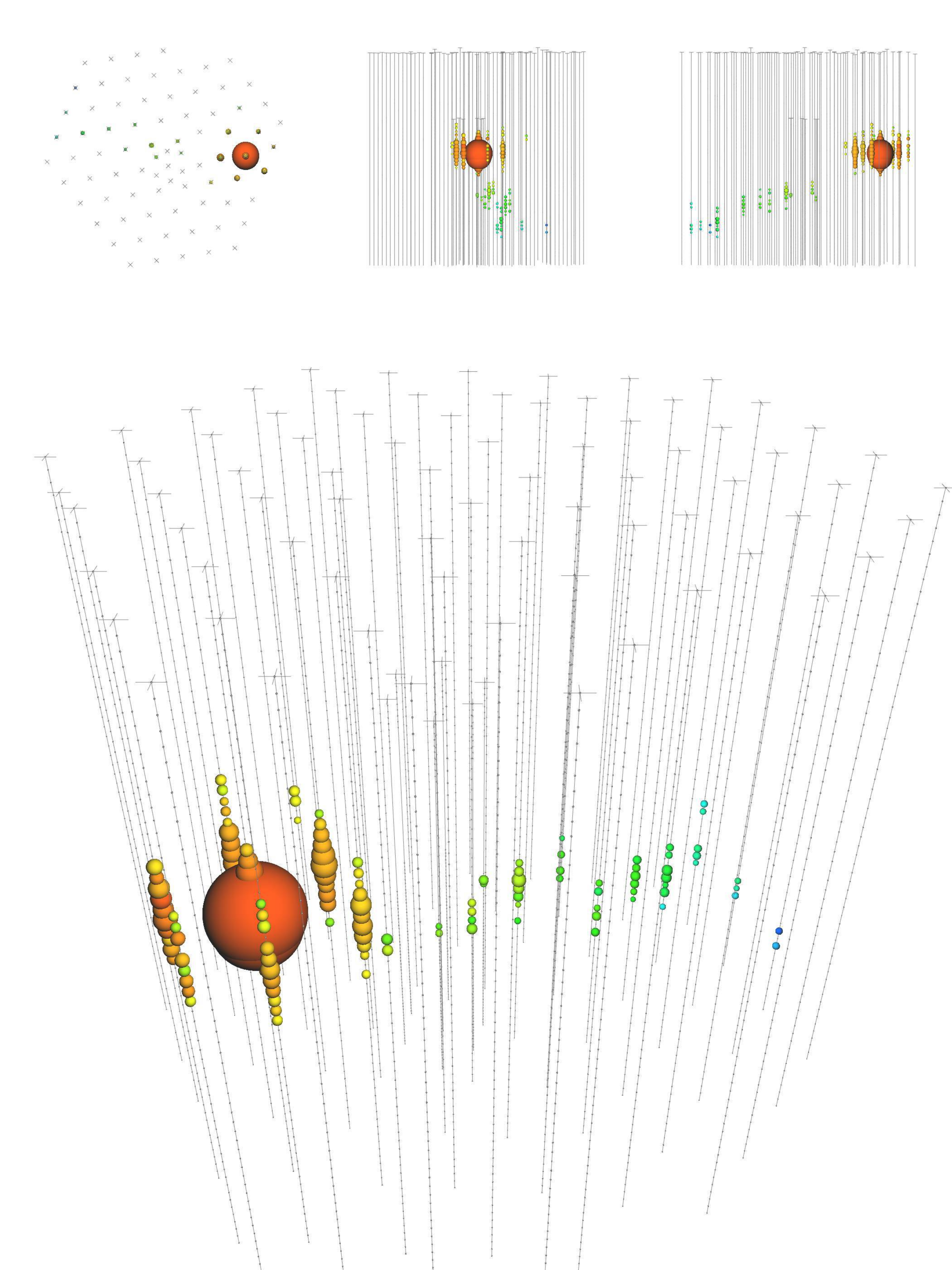}\\
\\
\includegraphics[width=0.8\linewidth]{color_scale.pdf}\vspace{0.2in}\\
\begin{tabular}{c|c|c|c|c|c}
Deposited Energy (TeV) & Time (MJD) & Declination (deg.) & RA (deg.) & Med. Ang. Resolution (deg.) & Topology\\
\hline
$32.6 \,^{+10.3}_{-11.1}$ & 55608.8201277 & $-21.2$ & $182.4$ & $\lesssim 1.3$ & Track
\end{tabular}
\newpage
\section*{Event 9}

\includegraphics[width=0.8\linewidth]{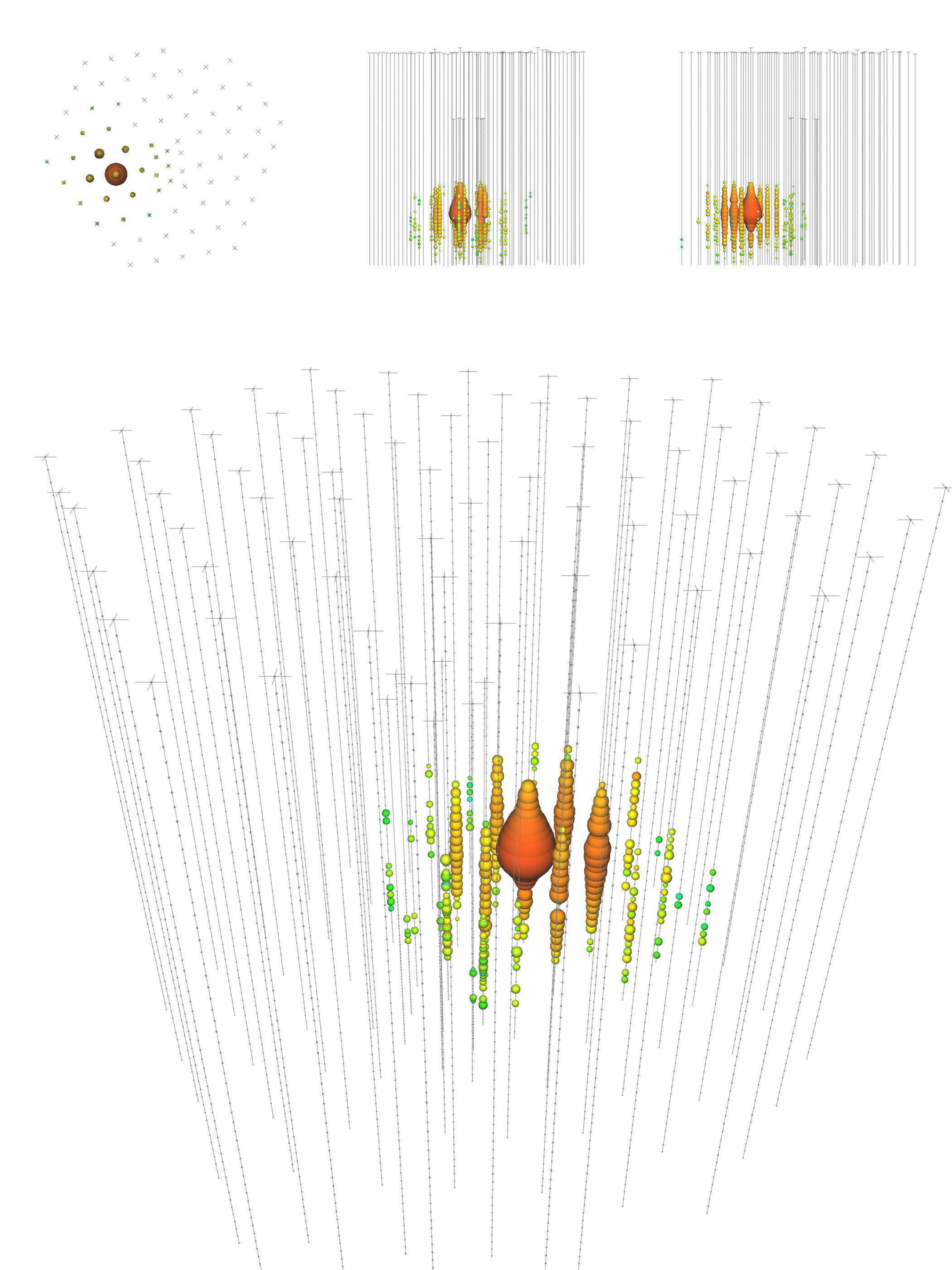}\\
\\
\includegraphics[width=0.8\linewidth]{color_scale.pdf}\vspace{0.2in}\\
\begin{tabular}{c|c|c|c|c|c}
Deposited Energy (TeV) & Time (MJD) & Declination (deg.) & RA (deg.) & Med. Ang. Resolution (deg.) & Topology\\
\hline
$63.2 \,^{+7.1}_{-8.0}$ & 55685.6629638 & $33.6$ & $151.3$ & $16.5$ & Shower
\end{tabular}
\newpage
\section*{Event 10}

\includegraphics[width=0.8\linewidth]{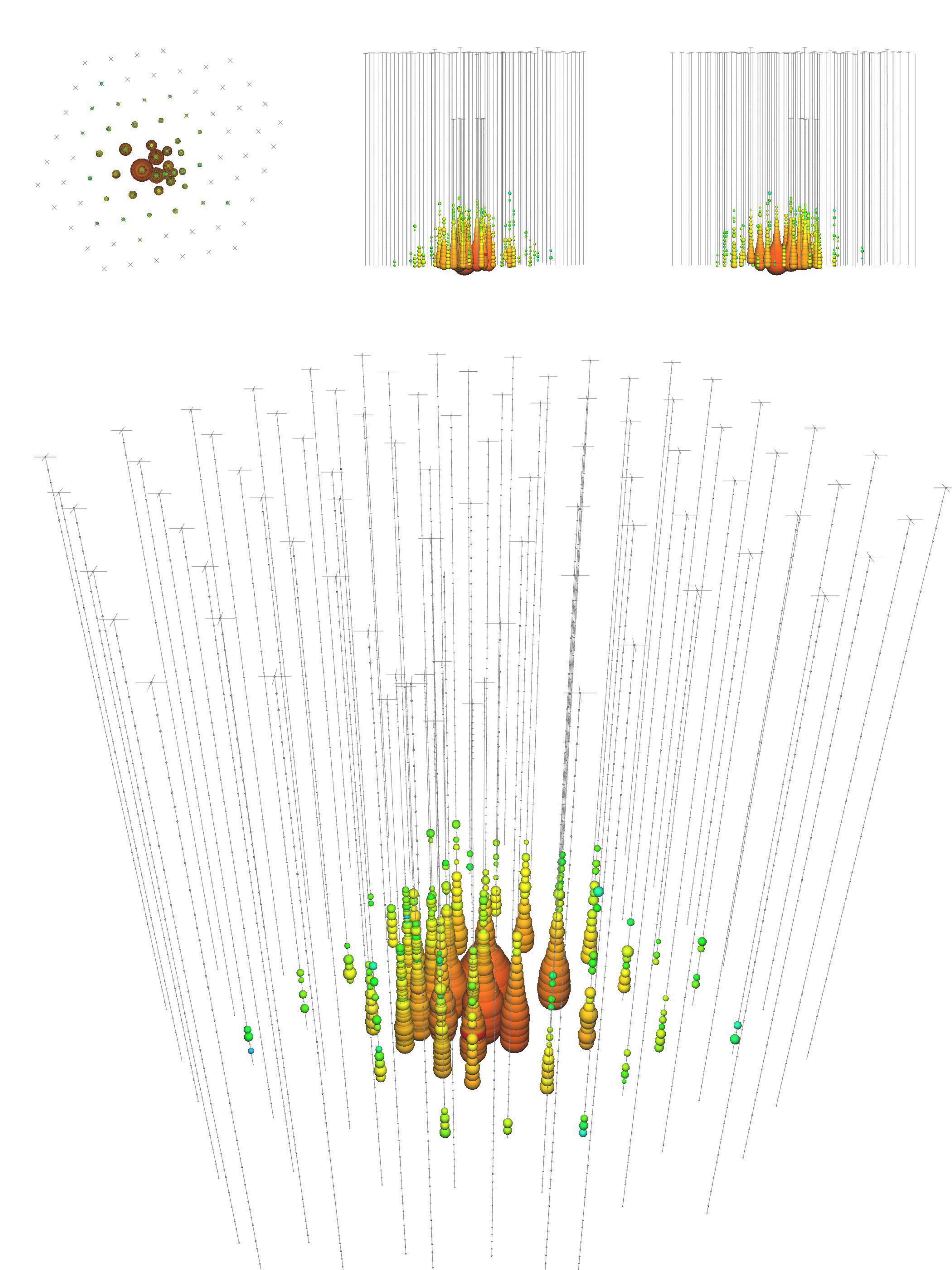}\\
\\
\includegraphics[width=0.8\linewidth]{color_scale.pdf}\vspace{0.2in}\\
\begin{tabular}{c|c|c|c|c|c}
Deposited Energy (TeV) & Time (MJD) & Declination (deg.) & RA (deg.) & Med. Ang. Resolution (deg.) & Topology\\
\hline
$97.2 \,^{+10.4}_{-12.4}$ & 55695.2730442 & $-29.4$ & $5.0$ & $8.1$ & Shower
\end{tabular}
\newpage
\section*{Event 11}

\includegraphics[width=0.8\linewidth]{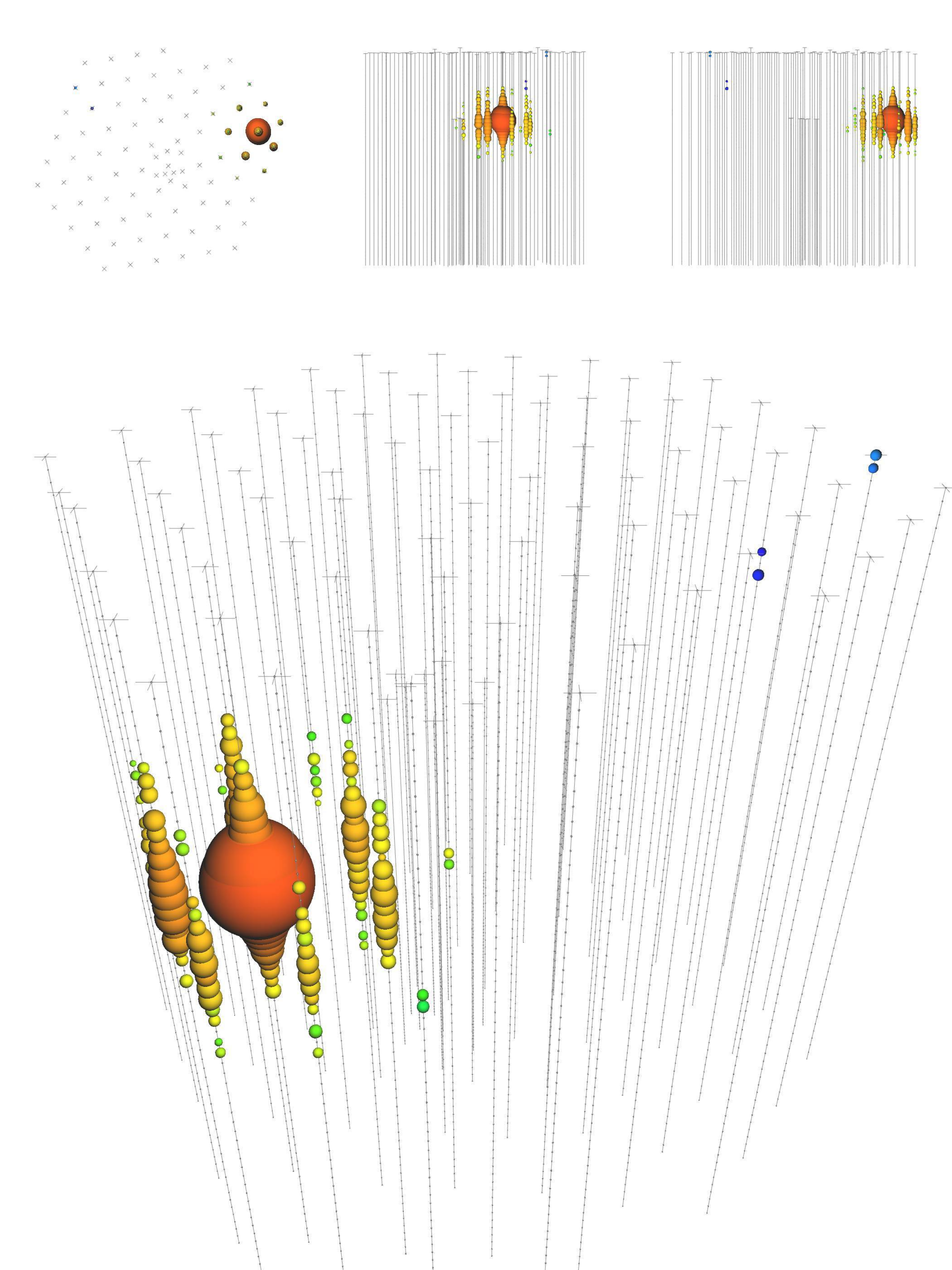}\\
\\
\includegraphics[width=0.8\linewidth]{color_scale.pdf}\vspace{0.2in}\\
\begin{tabular}{c|c|c|c|c|c}
Deposited Energy (TeV) & Time (MJD) & Declination (deg.) & RA (deg.) & Med. Ang. Resolution (deg.) & Topology\\
\hline
$88.4 \,^{+12.5}_{-10.7}$ & 55714.5909268 & $-8.9$ & $155.3$ & $16.7$ & Shower
\end{tabular}
\newpage
\section*{Event 12}

\includegraphics[width=0.8\linewidth]{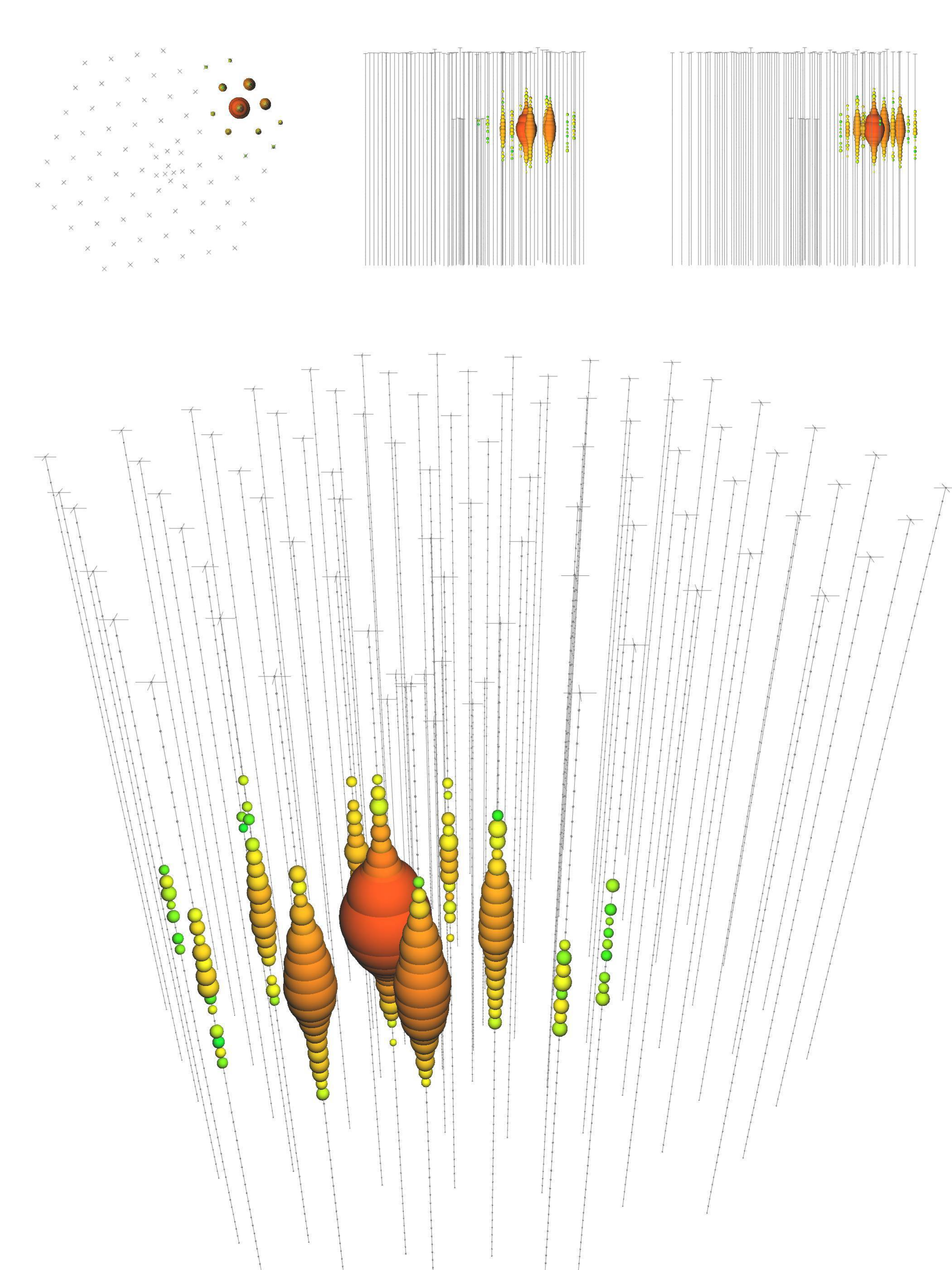}\\
\\
\includegraphics[width=0.8\linewidth]{color_scale.pdf}\vspace{0.2in}\\
\begin{tabular}{c|c|c|c|c|c}
Deposited Energy (TeV) & Time (MJD) & Declination (deg.) & RA (deg.) & Med. Ang. Resolution (deg.) & Topology\\
\hline
$104.1 \,^{+12.5}_{-13.2}$ & 55739.4411227 & $-52.8$ & $296.1$ & $9.8$ & Shower
\end{tabular}
\newpage
\section*{Event 13}

\includegraphics[width=0.8\linewidth]{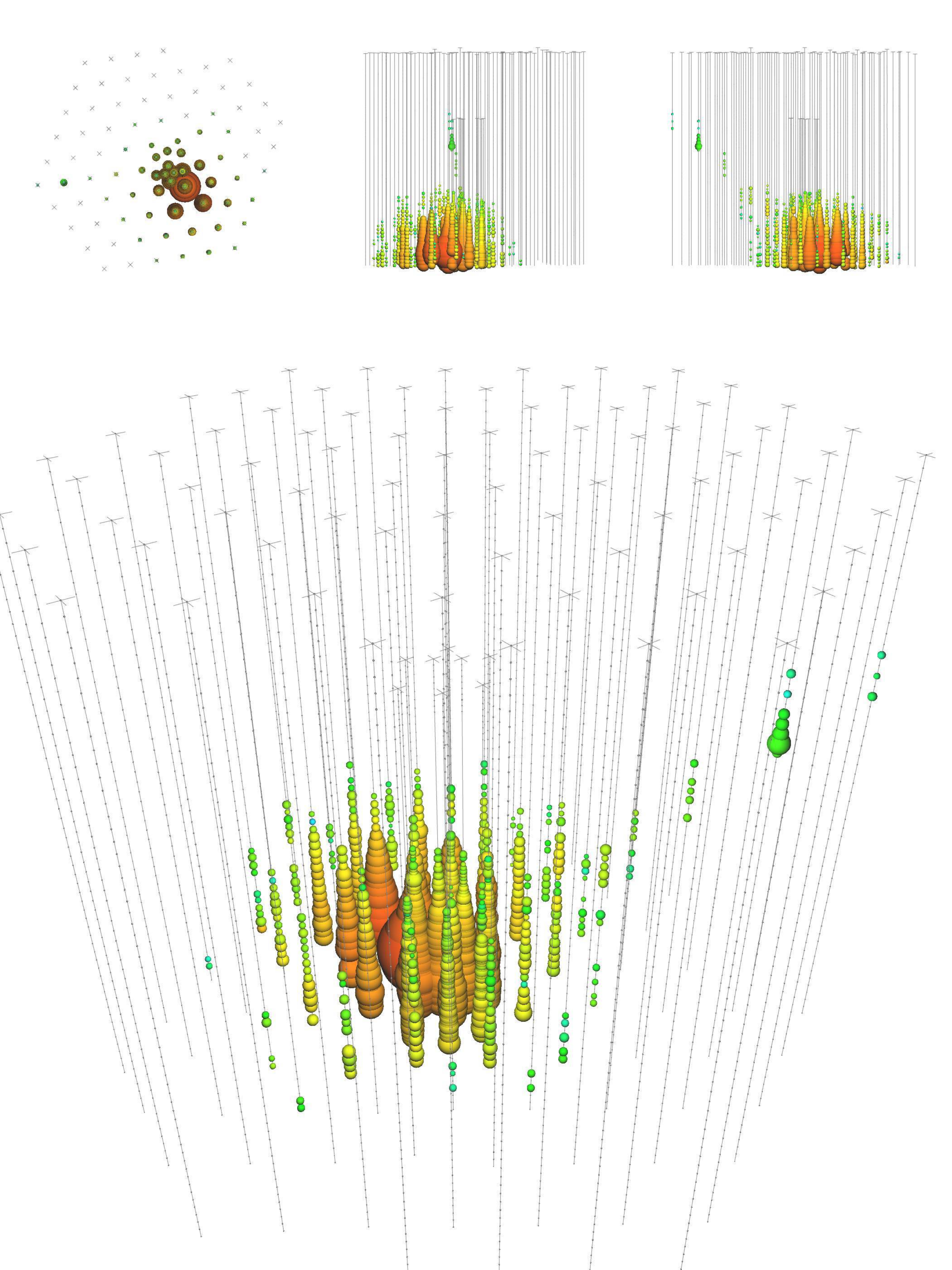}\\
\\
\includegraphics[width=0.8\linewidth]{color_scale.pdf}\vspace{0.2in}\\
\begin{tabular}{c|c|c|c|c|c}
Deposited Energy (TeV) & Time (MJD) & Declination (deg.) & RA (deg.) & Med. Ang. Resolution (deg.) & Topology\\
\hline
$252.7 \,^{+25.9}_{-21.6}$ & 55756.1129755 & $40.3$ & $67.9$ & $\lesssim 1.2$ & Track
\end{tabular}
\newpage
\section*{Event 14}

\includegraphics[width=0.8\linewidth]{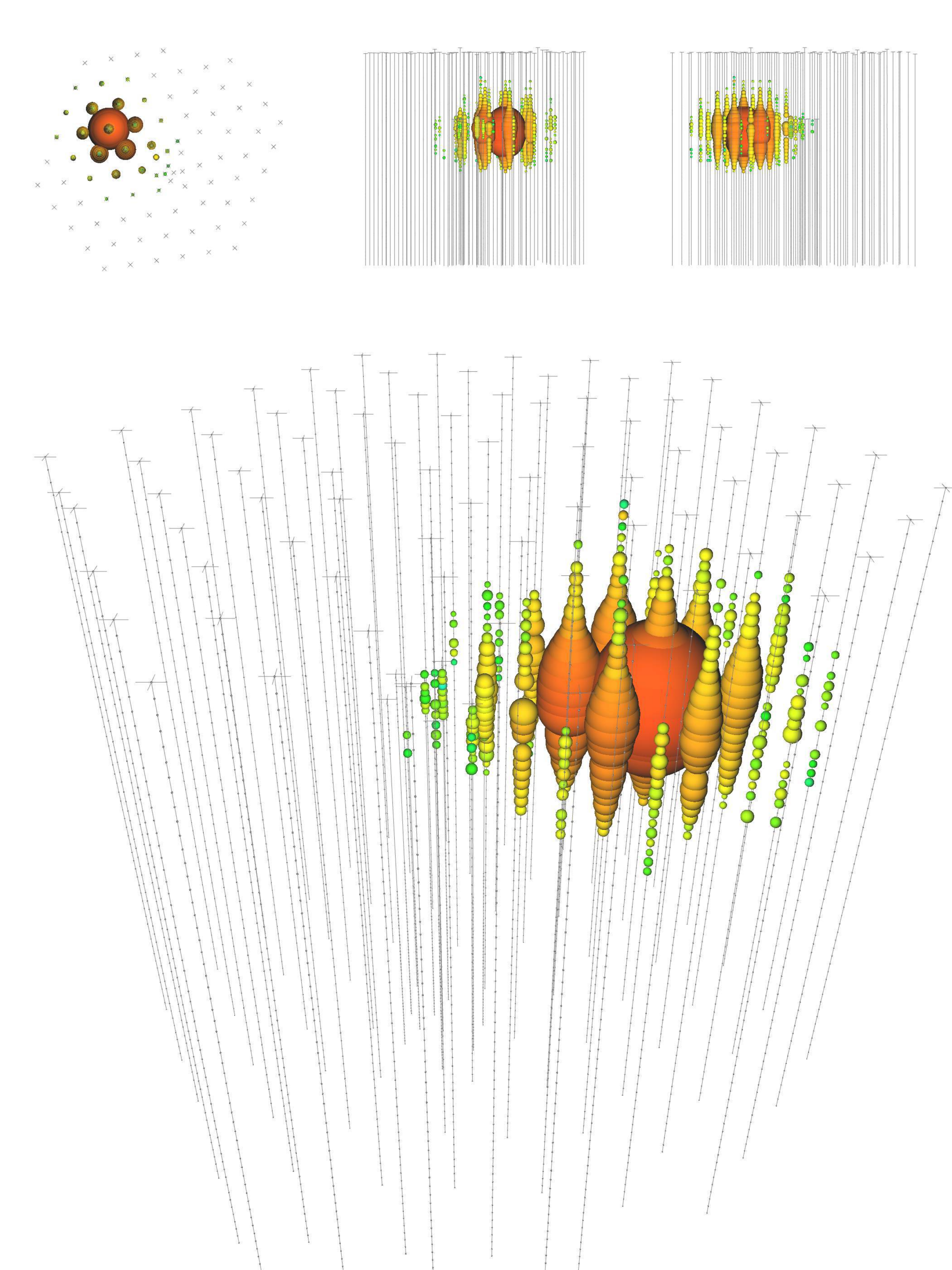}\\
\\
\includegraphics[width=0.8\linewidth]{color_scale.pdf}\vspace{0.2in}\\
\begin{tabular}{c|c|c|c|c|c}
Deposited Energy (TeV) & Time (MJD) & Declination (deg.) & RA (deg.) & Med. Ang. Resolution (deg.) & Topology\\
\hline
$1040.7 \,^{+131.6}_{-144.4}$ & 55782.5161816 & $-27.9$ & $265.6$ & $13.2$ & Shower
\end{tabular}
\newpage
\section*{Event 15}

\includegraphics[width=0.8\linewidth]{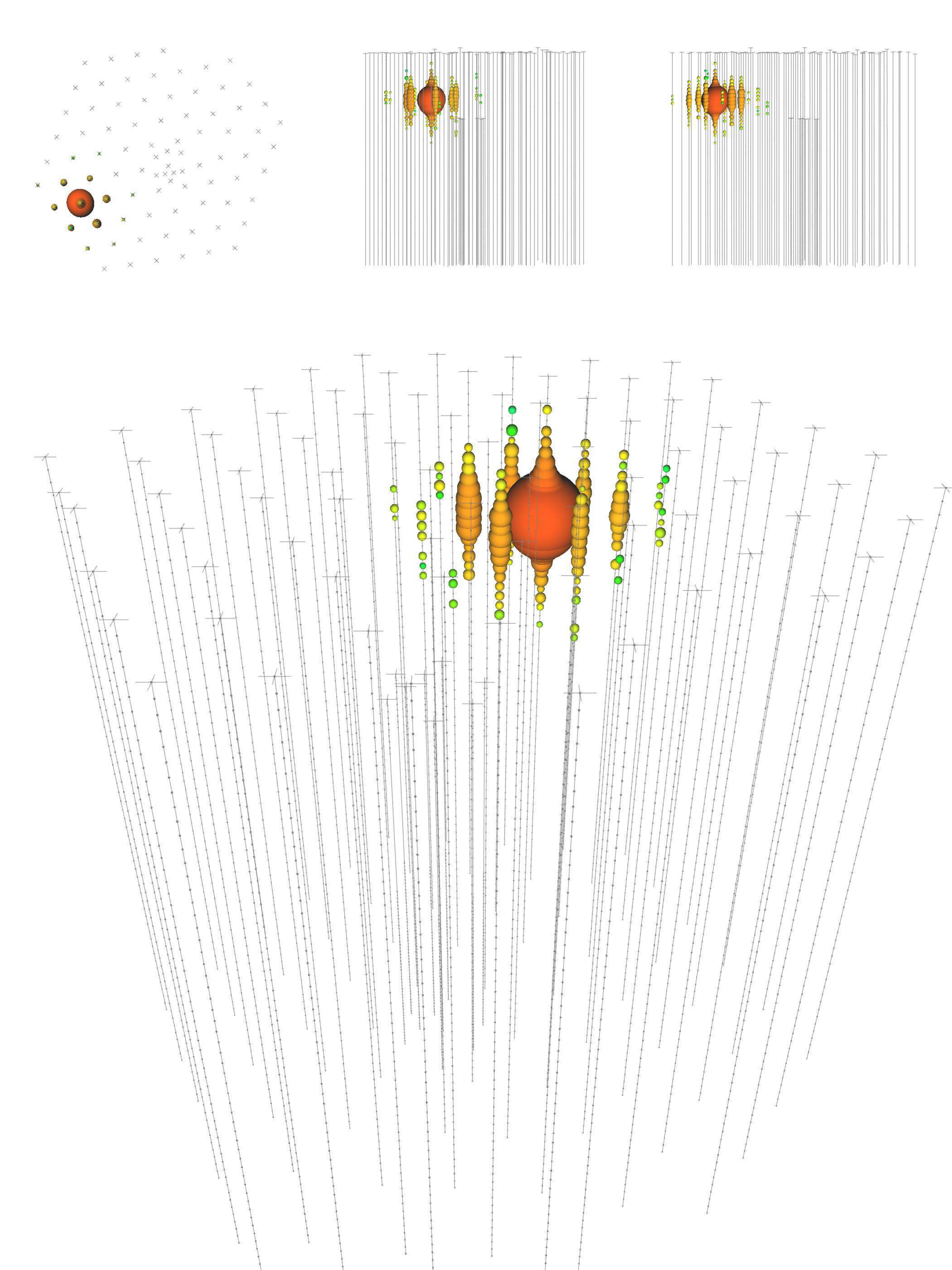}\\
\\
\includegraphics[width=0.8\linewidth]{color_scale.pdf}\vspace{0.2in}\\
\begin{tabular}{c|c|c|c|c|c}
Deposited Energy (TeV) & Time (MJD) & Declination (deg.) & RA (deg.) & Med. Ang. Resolution (deg.) & Topology\\
\hline
$57.5 \,^{+8.3}_{-7.8}$ & 55783.1854172 & $-49.7$ & $287.3$ & $19.7$ & Shower
\end{tabular}
\newpage
\section*{Event 16}

\includegraphics[width=0.8\linewidth]{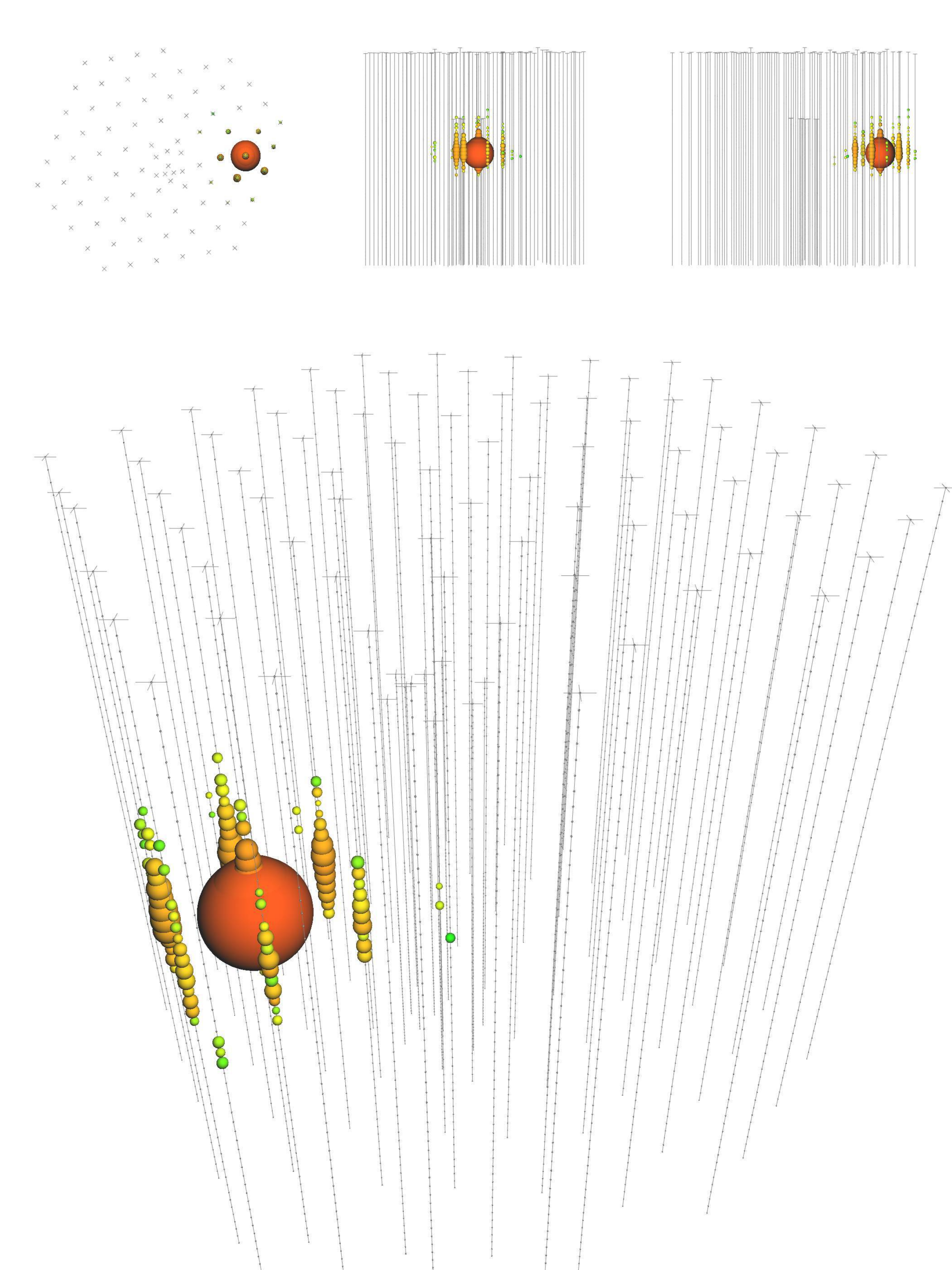}\\
\\
\includegraphics[width=0.8\linewidth]{color_scale.pdf}\vspace{0.2in}\\
\begin{tabular}{c|c|c|c|c|c}
Deposited Energy (TeV) & Time (MJD) & Declination (deg.) & RA (deg.) & Med. Ang. Resolution (deg.) & Topology\\
\hline
$30.6 \,^{+3.6}_{-3.5}$ & 55798.6271191 & $-22.6$ & $192.1$ & $19.4$ & Shower
\end{tabular}
\newpage
\section*{Event 17}

\includegraphics[width=0.8\linewidth]{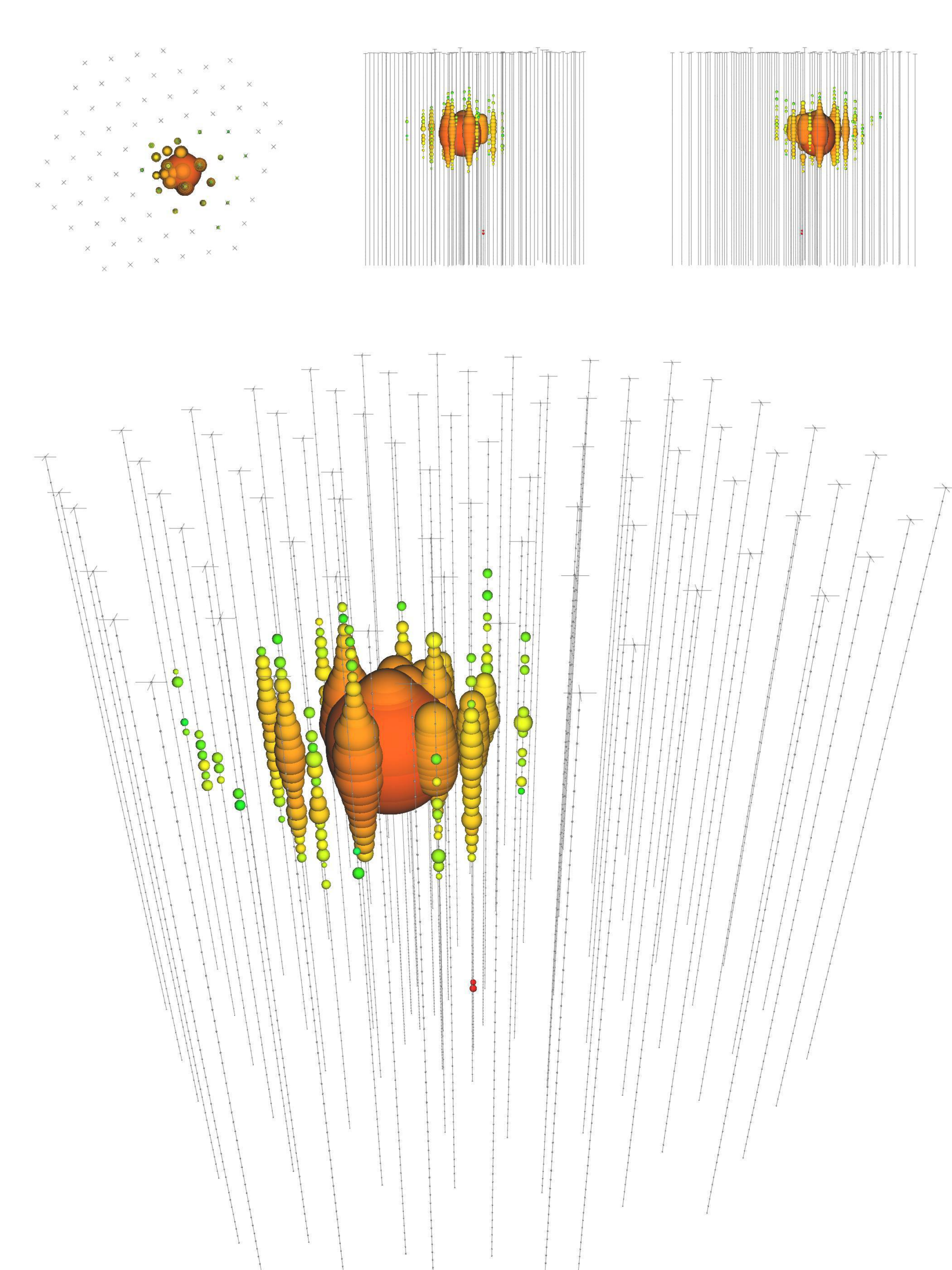}\\
\\
\includegraphics[width=0.8\linewidth]{color_scale.pdf}\vspace{0.2in}\\
\begin{tabular}{c|c|c|c|c|c}
Deposited Energy (TeV) & Time (MJD) & Declination (deg.) & RA (deg.) & Med. Ang. Resolution (deg.) & Topology\\
\hline
$199.7 \,^{+27.2}_{-26.8}$ & 55800.3755444 & $14.5$ & $247.4$ & $11.6$ & Shower
\end{tabular}
\newpage
\section*{Event 18}

\includegraphics[width=0.8\linewidth]{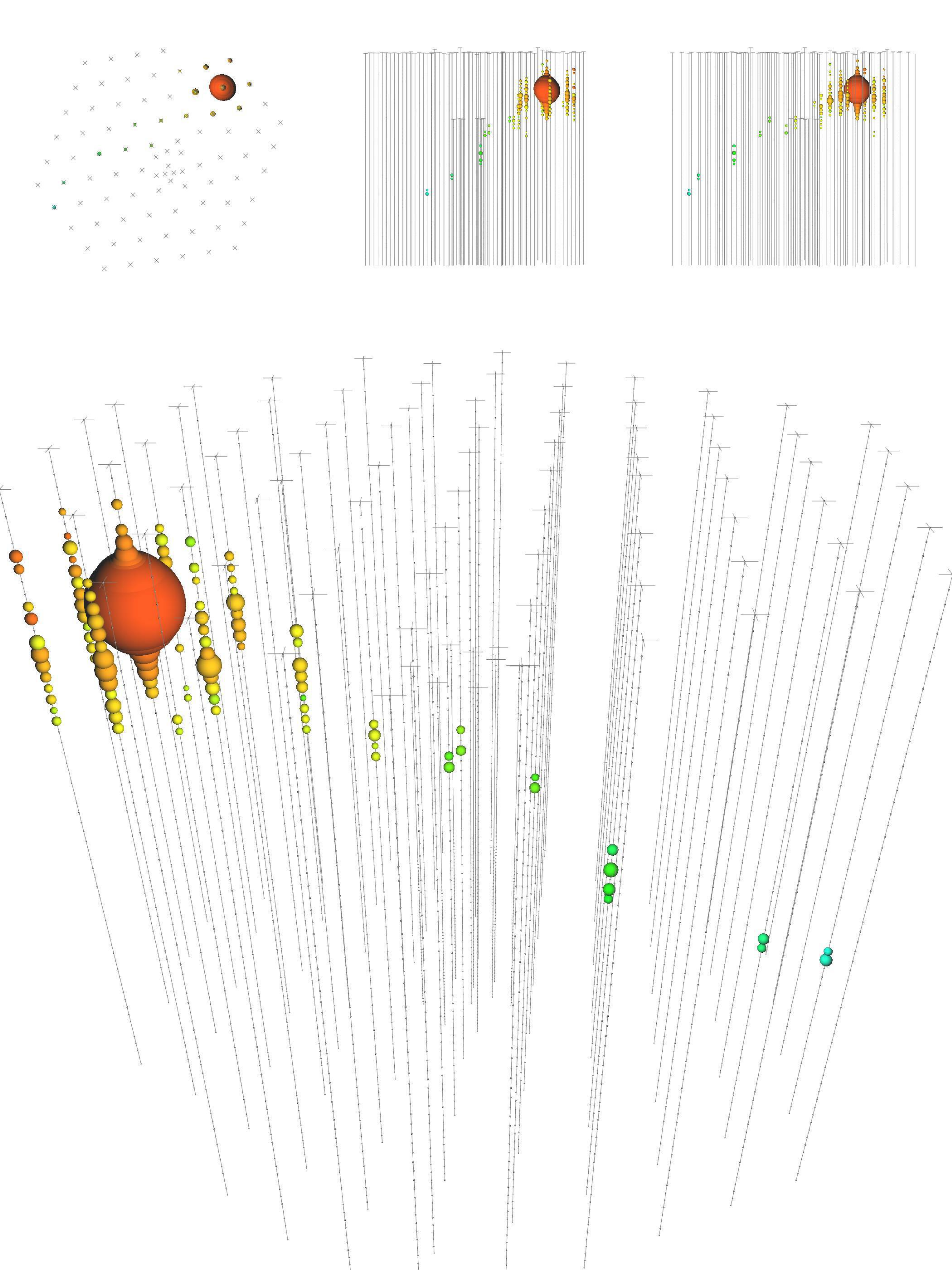}\\
\\
\includegraphics[width=0.8\linewidth]{color_scale.pdf}\vspace{0.2in}\\
\begin{tabular}{c|c|c|c|c|c}
Deposited Energy (TeV) & Time (MJD) & Declination (deg.) & RA (deg.) & Med. Ang. Resolution (deg.) & Topology\\
\hline
$31.5 \,^{+4.6}_{-3.3}$ & 55923.5318175 & $-24.8$ & $345.6$ & $\lesssim 1.3$ & Track
\end{tabular}
\newpage
\section*{Event 19}

\includegraphics[width=0.8\linewidth]{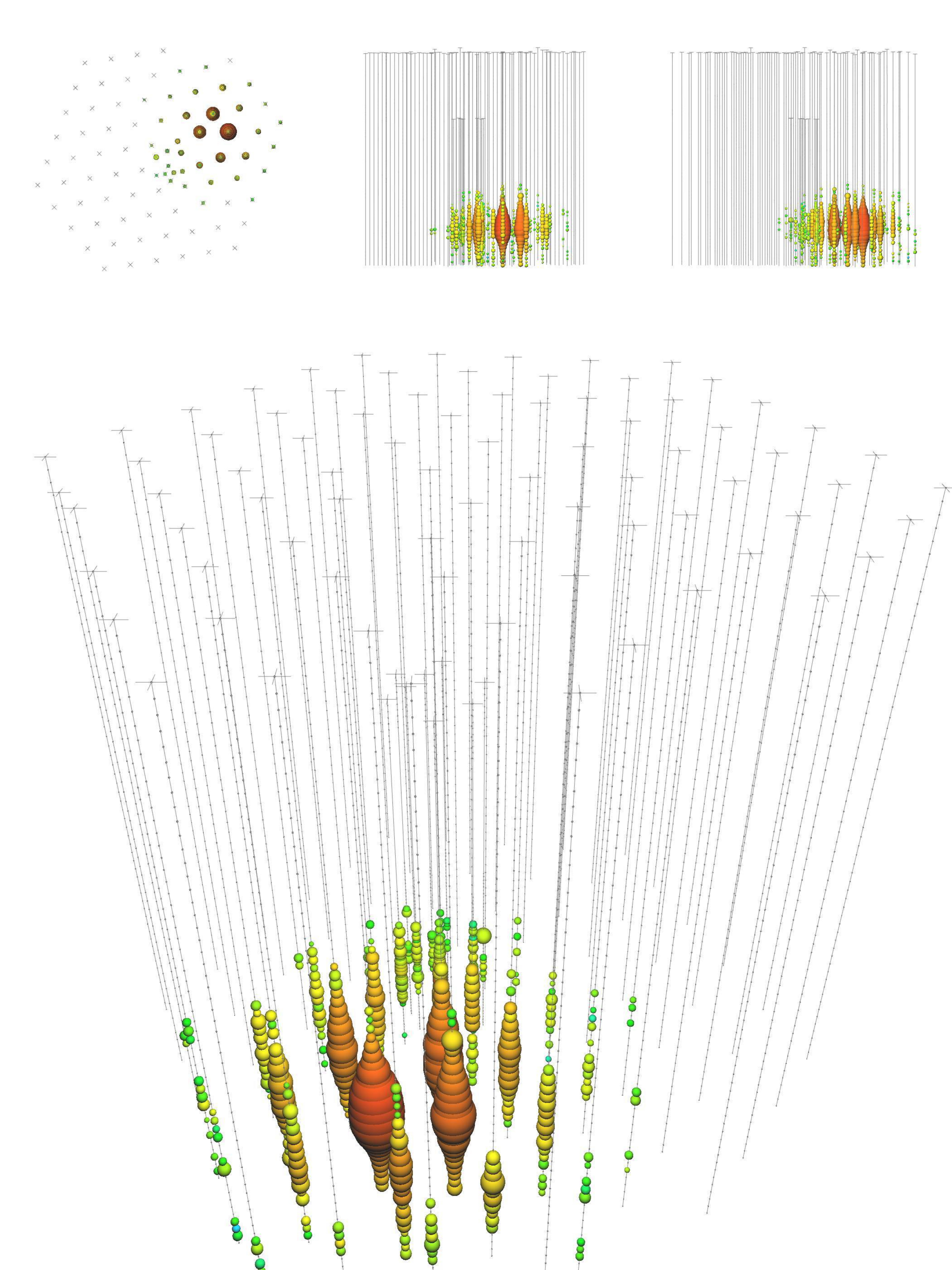}\\
\\
\includegraphics[width=0.8\linewidth]{color_scale.pdf}\vspace{0.2in}\\
\begin{tabular}{c|c|c|c|c|c}
Deposited Energy (TeV) & Time (MJD) & Declination (deg.) & RA (deg.) & Med. Ang. Resolution (deg.) & Topology\\
\hline
$71.5 \,^{+7.0}_{-7.2}$ & 55925.7958570 & $-59.7$ & $76.9$ & $9.7$ & Shower
\end{tabular}
\newpage
\section*{Event 20}

\includegraphics[width=0.8\linewidth]{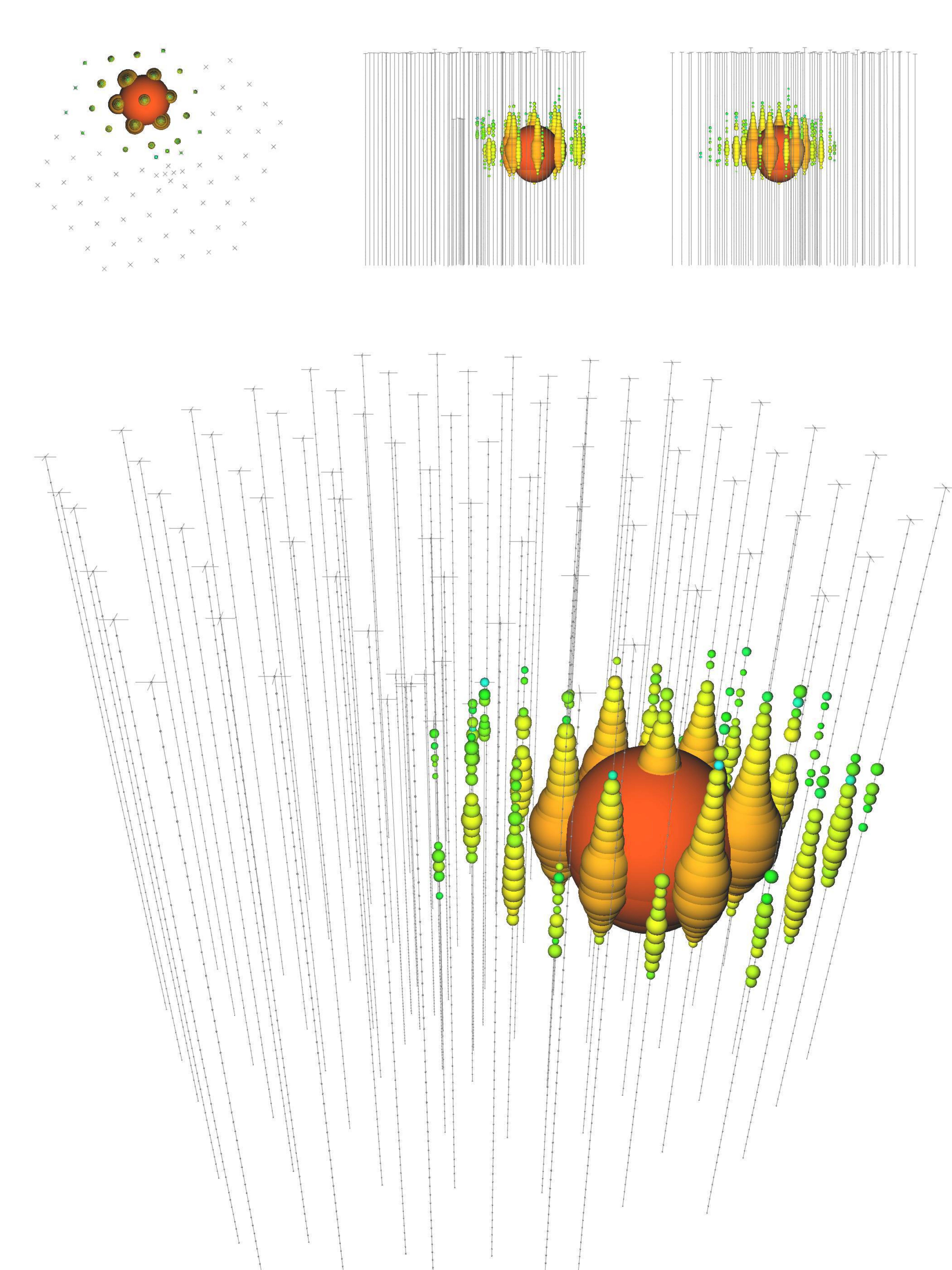}\\
\\
\includegraphics[width=0.8\linewidth]{color_scale.pdf}\vspace{0.2in}\\
\begin{tabular}{c|c|c|c|c|c}
Deposited Energy (TeV) & Time (MJD) & Declination (deg.) & RA (deg.) & Med. Ang. Resolution (deg.) & Topology\\
\hline
$1140.8 \,^{+142.8}_{-132.8}$ & 55929.3986232 & $-67.2$ & $38.3$ & $10.7$ & Shower
\end{tabular}
\newpage
\section*{Event 21}

\includegraphics[width=0.8\linewidth]{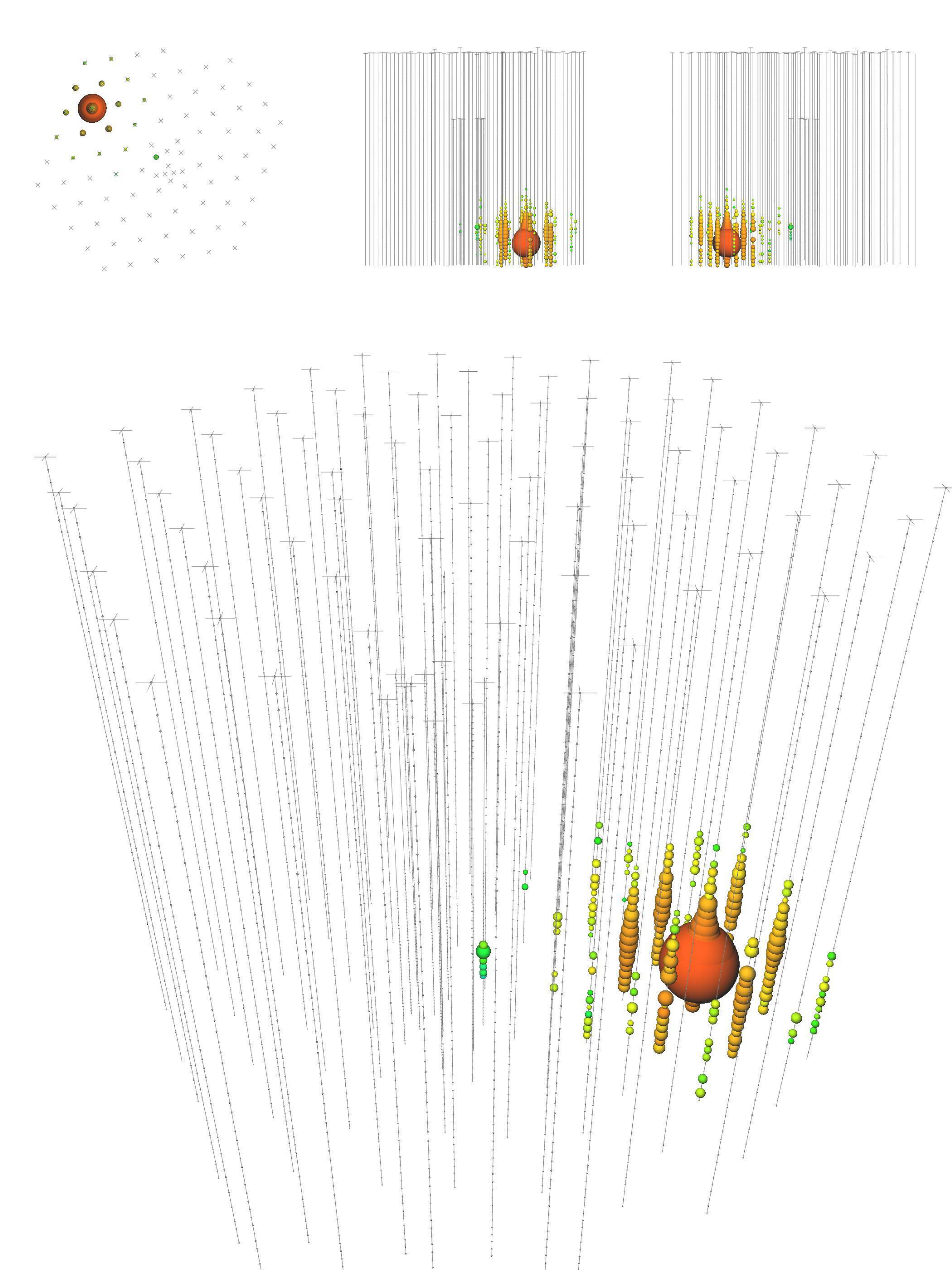}\\
\\
\includegraphics[width=0.8\linewidth]{color_scale.pdf}\vspace{0.2in}\\
\begin{tabular}{c|c|c|c|c|c}
Deposited Energy (TeV) & Time (MJD) & Declination (deg.) & RA (deg.) & Med. Ang. Resolution (deg.) & Topology\\
\hline
$30.2 \,^{+3.5}_{-3.3}$ & 55936.5416440 & $-24.0$ & $9.0$ & $20.9$ & Shower
\end{tabular}
\newpage
\section*{Event 22}

\includegraphics[width=0.8\linewidth]{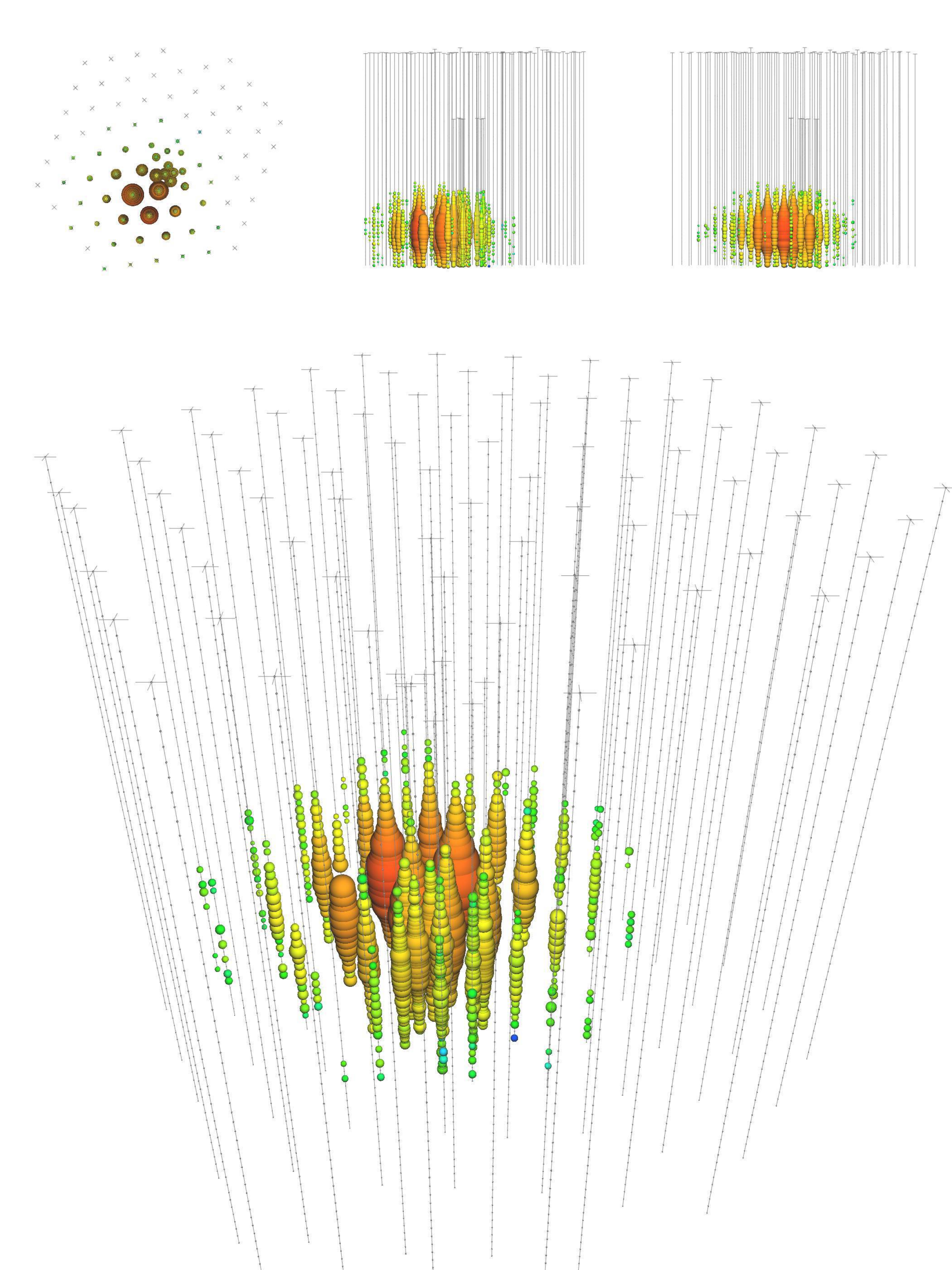}\\
\\
\includegraphics[width=0.8\linewidth]{color_scale.pdf}\vspace{0.2in}\\
\begin{tabular}{c|c|c|c|c|c}
Deposited Energy (TeV) & Time (MJD) & Declination (deg.) & RA (deg.) & Med. Ang. Resolution (deg.) & Topology\\
\hline
$219.5 \,^{+21.2}_{-24.4}$ & 55941.9757760 & $-22.1$ & $293.7$ & $12.1$ & Shower
\end{tabular}
\newpage
\section*{Event 23}

\includegraphics[width=0.8\linewidth]{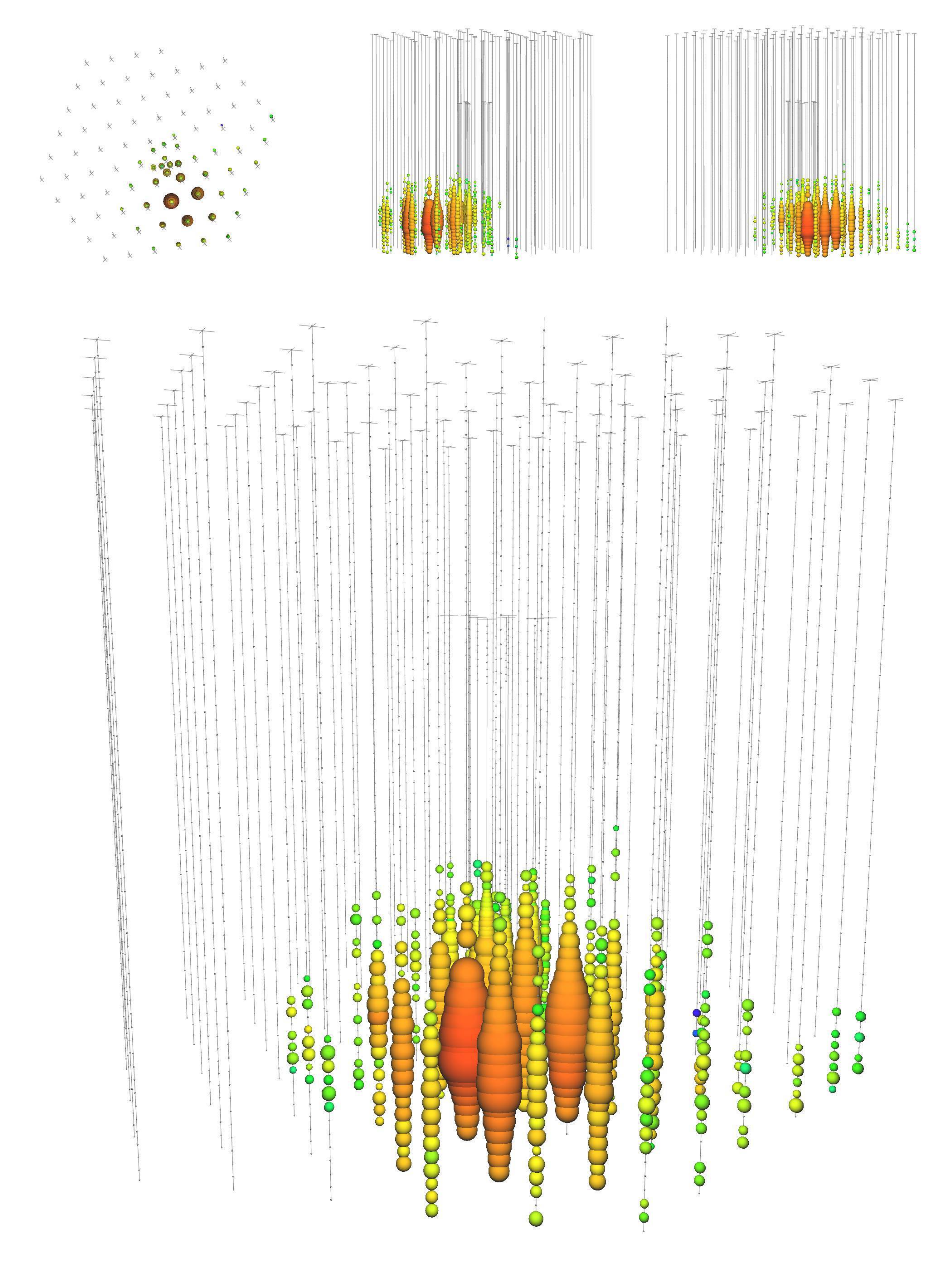}\\
\\
\includegraphics[width=0.8\linewidth]{color_scale.pdf}\vspace{0.2in}\\
\begin{tabular}{c|c|c|c|c|c}
Deposited Energy (TeV) & Time (MJD) & Declination (deg.) & RA (deg.) & Med. Ang. Resolution (deg.) & Topology\\
\hline
$82.2 \,^{+8.6}_{-8.4}$ & 55949.5693177 & $-13.2$ & $208.7$ & $\lesssim 1.9$ & Track
\end{tabular}
\newpage
\section*{Event 24}

\includegraphics[width=0.8\linewidth]{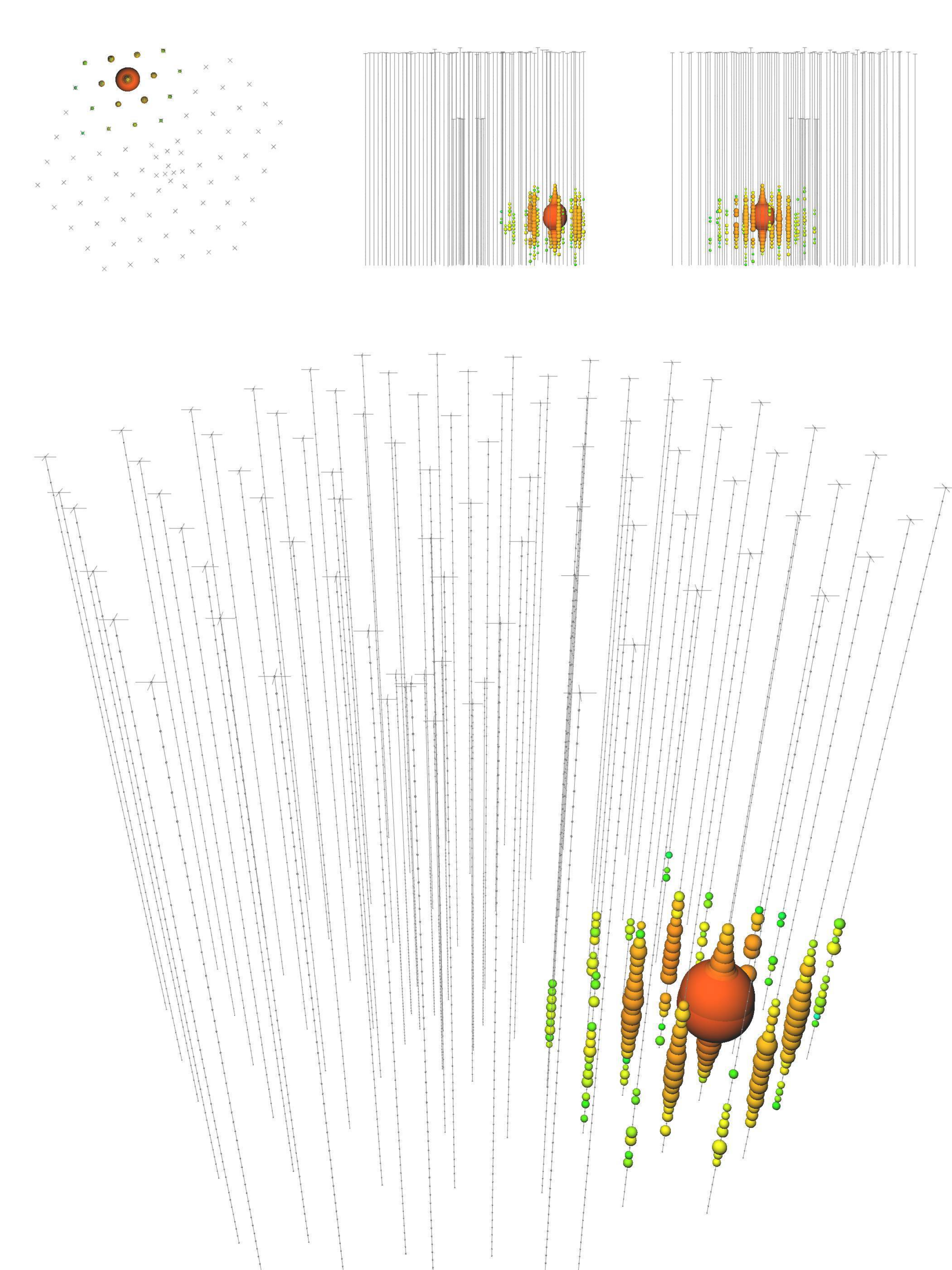}\\
\\
\includegraphics[width=0.8\linewidth]{color_scale.pdf}\vspace{0.2in}\\
\begin{tabular}{c|c|c|c|c|c}
Deposited Energy (TeV) & Time (MJD) & Declination (deg.) & RA (deg.) & Med. Ang. Resolution (deg.) & Topology\\
\hline
$30.5 \,^{+3.2}_{-2.6}$ & 55950.8474887 & $-15.1$ & $282.2$ & $15.5$ & Shower
\end{tabular}
\newpage
\section*{Event 25}

\includegraphics[width=0.8\linewidth]{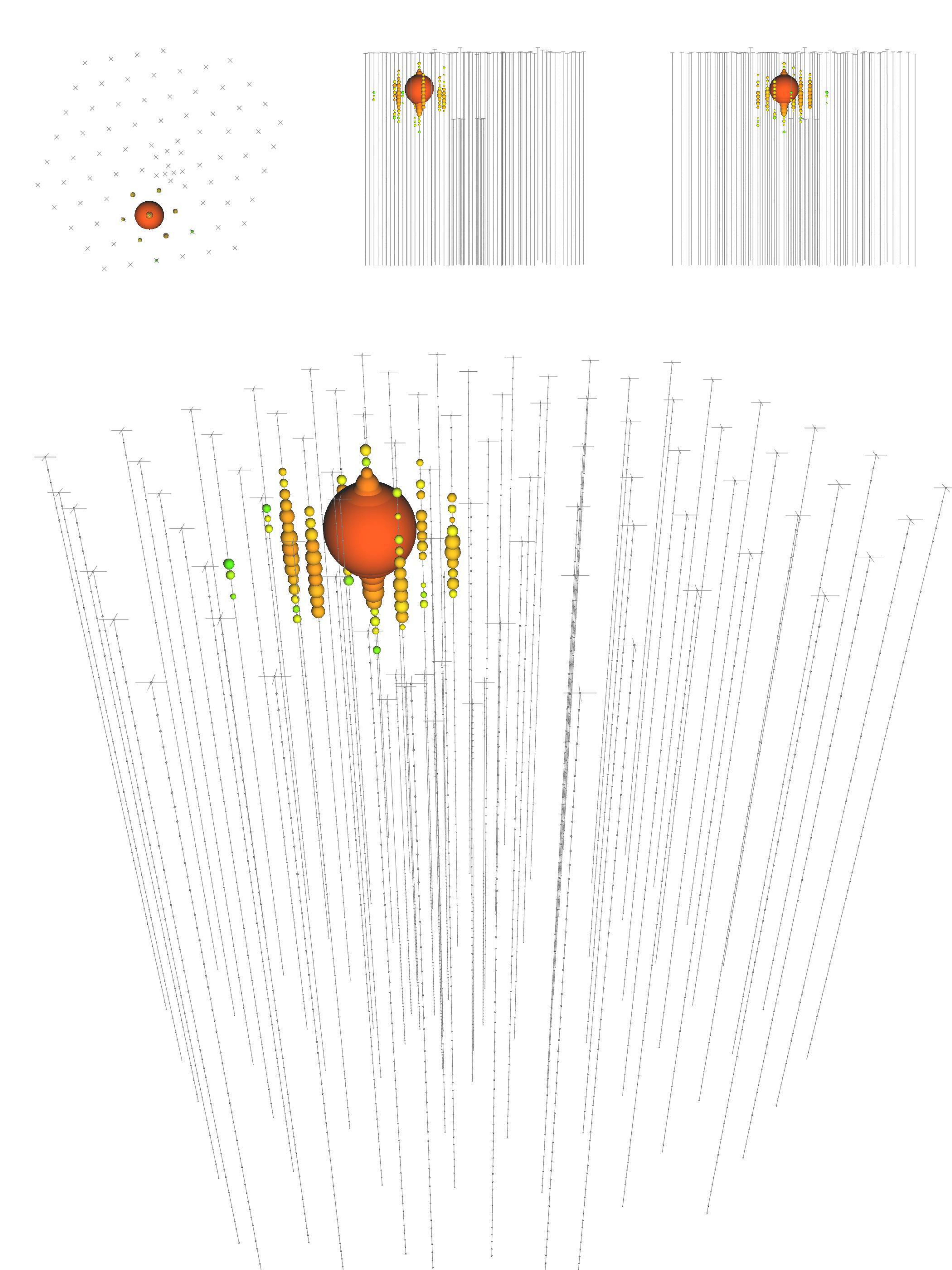}\\
\\
\includegraphics[width=0.8\linewidth]{color_scale.pdf}\vspace{0.2in}\\
\begin{tabular}{c|c|c|c|c|c}
Deposited Energy (TeV) & Time (MJD) & Declination (deg.) & RA (deg.) & Med. Ang. Resolution (deg.) & Topology\\
\hline
$33.5 \,^{+4.9}_{-5.0}$ & 55966.7422457 & $-14.5$ & $286.0$ & $46.3$ & Shower
\end{tabular}
\newpage
\section*{Event 26}

\includegraphics[width=0.8\linewidth]{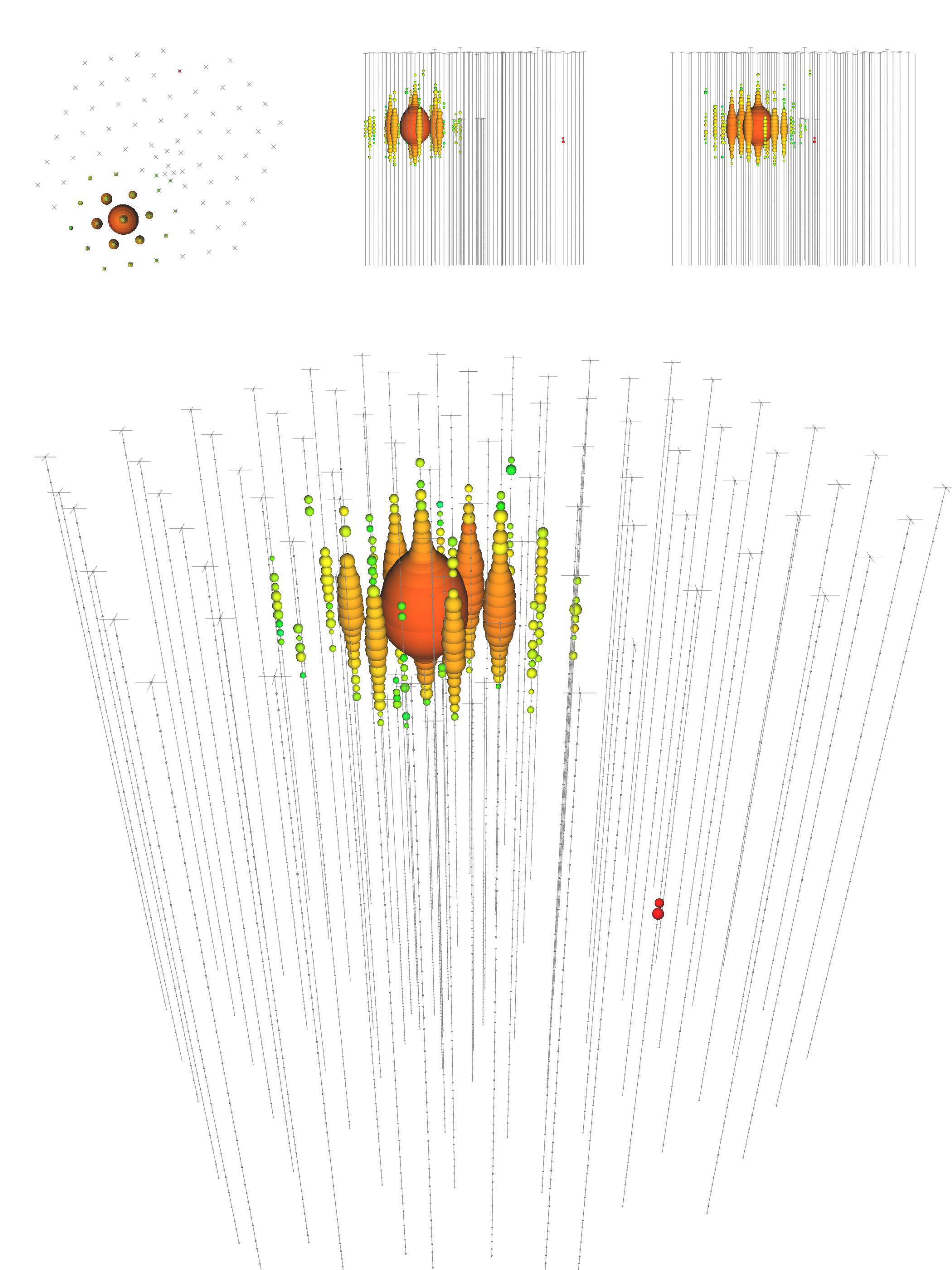}\\
\\
\includegraphics[width=0.8\linewidth]{color_scale.pdf}\vspace{0.2in}\\
\begin{tabular}{c|c|c|c|c|c}
Deposited Energy (TeV) & Time (MJD) & Declination (deg.) & RA (deg.) & Med. Ang. Resolution (deg.) & Topology\\
\hline
$210.0 \,^{+29.0}_{-25.8}$ & 55979.2551738 & $22.7$ & $143.4$ & $11.8$ & Shower
\end{tabular}
\newpage
\section*{Event 27}

\includegraphics[width=0.8\linewidth]{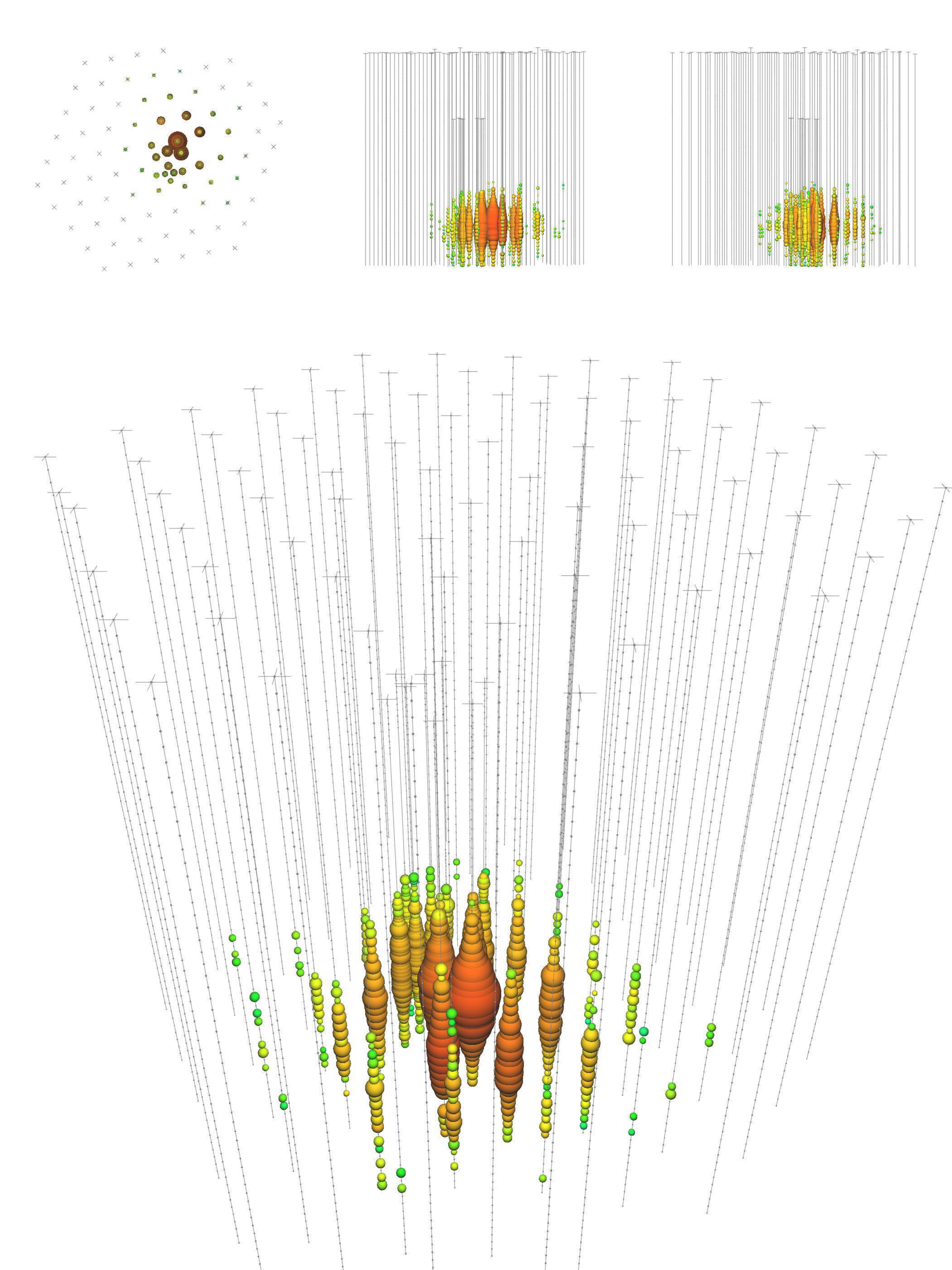}\\
\\
\includegraphics[width=0.8\linewidth]{color_scale.pdf}\vspace{0.2in}\\
\begin{tabular}{c|c|c|c|c|c}
Deposited Energy (TeV) & Time (MJD) & Declination (deg.) & RA (deg.) & Med. Ang. Resolution (deg.) & Topology\\
\hline
$60.2 \,^{+5.6}_{-5.6}$ & 56008.6845606 & $-12.6$ & $121.7$ & $6.6$ & Shower
\end{tabular}
\newpage
\section*{Event 28}

\includegraphics[width=0.8\linewidth]{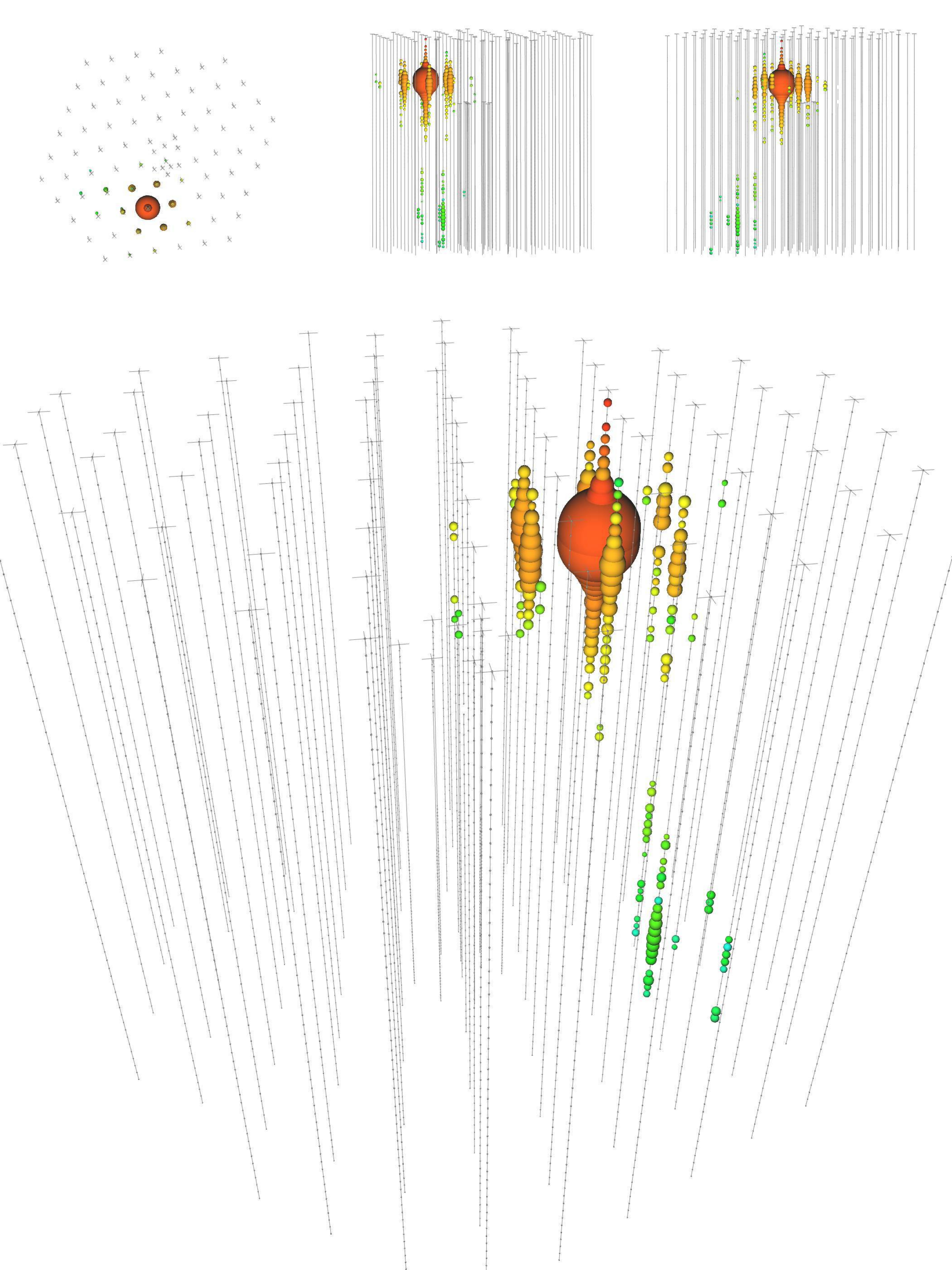}\\
\\
\includegraphics[width=0.8\linewidth]{color_scale.pdf}\vspace{0.2in}\\
\begin{tabular}{c|c|c|c|c|c}
Deposited Energy (TeV) & Time (MJD) & Declination (deg.) & RA (deg.) & Med. Ang. Resolution (deg.) & Topology\\
\hline
$46.1 \,^{+5.7}_{-4.4}$ & 56048.5704171 & $-71.5$ & $164.8$ & $\lesssim 1.3$ & Track
\end{tabular}
\fi

\end{document}